\documentstyle[prl,epsf,floats,aps,amsfonts,twocolumn]{revtex}

\input{epsf.tex}

\newcommand {\be}{\begin{equation}}
\newcommand {\ee} {\end{equation}}
\newcommand {\ba}{\begin{eqnarray}}
\newcommand {\ea} {\end{eqnarray}}

\def \F2 {FPL${}^2$ }

\def \OMIT #1{}
\def \rem #1 {{\it #1}}

\begin{document}

\twocolumn[\hsize\textwidth\columnwidth\hsize\csname
@twocolumnfalse\endcsname

\title {Minimum spanning trees and random resistor networks in $d$ dimensions}
\author {N. Read}

\address{Department of Physics, Yale University, P.O. Box
208120, New Haven, CT 06520-8120, USA}

\date{April 11, 2005}
\maketitle

\begin{abstract}
We consider minimum-cost spanning trees, both in lattice and
Euclidean models, in $d$ dimensions. For the cost of the optimum
tree in a box of size $L$, we show that there is a correction of
order $L^\theta$, where $\theta\leq 0$ is a universal
$d$-dependent exponent. There is a similar form for the change in
optimum cost under a change in boundary condition. At non-zero
temperature $T$, there is a crossover length $\xi\sim T^{-\nu}$,
such that on length scales larger than $\xi$, the behavior becomes
that of {\em uniform} spanning trees. There is a scaling relation
$\theta=-1/\nu$, and we provide several arguments that show that
$\nu$ and $-1/\theta$ both equal $\nu_{\rm perc}$, the correlation
length exponent for ordinary percolation in the same dimension
$d$, in all dimensions $d\geq 1$. The arguments all rely on the
close relation of Kruskal's greedy algorithm for the minimum
spanning tree, percolation, and (for some arguments) random
resistor networks. The scaling of the entropy and free energy at
small non-zero $T$, and hence of the number of near-optimal
solutions, is also discussed. We suggest that the Steiner tree
problem is in the same universality class as the minimum spanning
tree in all dimensions, as is the traveling salesman problem in
two dimensions. Hence all will have the same value of
$\theta=-3/4$ in two dimensions.
\end{abstract}
]


\section{Introduction}

Minimum spanning trees are a problem of combinatorial optimization
\cite{papst,steele}. Suppose we are given an undirected connected
graph $G$, with vertex set $V$ and edge set $E$, and a cost (or
weight, or ``length'') $\ell_{ij}$ assigned to each edge $\langle
ij\rangle \in E$ (where $i$, $j\in V$). The problem is to find a
spanning tree $\bf T$ (i.e.\ a connected subgraph of $G$ that
includes all vertices in $V$, but whose edges form no cycles; such
a tree must have $|V|-1$ edges),
such that the total cost of the edges in $\bf T$,%
\be%
\ell=\sum_{\langle ij \rangle \in \bf T} \ell_{ij} \ee%
is as small as possible. Thus the minimization is over the set
$\cal T$ of spanning trees in $G$.

In this paper we are interested in the case in which $G$ is a
simply-connected portion $\Lambda$ of a regular lattice in $d\geq
1$ dimensions (with edges connecting nearest-neighbor lattice
vertices only; the nearest-neighbor distance is fixed at $1$
throughout this paper), including the case when $\Lambda$ tends to
the entire lattice, and the edge costs are independent,
identically-distributed random variables, for example $\ell_{ij}$
uniformly distributed on $[0,1]$. We will also consider geometries
with periodic boundary conditions, in which $\Lambda$ has no
boundary. The results also apply without significant modification
to cases with other distributions, and/or with short-range
correlations of the $\ell_{ij}$s, and to the Euclidean minimum
spanning tree, in which $N=|V|$ points are distributed
independently and uniformly (with density 1) in a portion
$\Lambda$ of $d$-dimensional Euclidean space, and the cost of an
edge $\langle ij\rangle$ is the Euclidean distance between $i$ and
$j$, for any pair $i\neq j$.

The motivation for this work is to understand disordered systems
at low temperatures better, beginning with those in which
quantum-mechanical effects are negligible. Here ``disordered''
means that the Hamiltonian (or energy as a function of the system
configuration) contains random variables, and the minimum energy
must be found for fixed (or ``quenched'') values of these random
variables. Such systems include classical Ising spin glass models.
There is a great deal of overlap between this field and that of
random optimization, including some common models \cite{mpv}.
There is even a strongly-disordered spin-glass model that maps
onto minimum spanning trees \cite{ns}. The results in this paper
can be considered as a rare case in which some exact results (or
exact mapping to another problem) can be found for a fairly
natural system with quenched disorder.

The questions of interest here include the dependence of the total
cost of the minimum spanning tree (MST) on the size of the system
$\Lambda$, and on certain changes of boundary conditions to be
defined below. The expectation value of the cost $\ell_{\rm OPT}$
of the MST is expected to take the form (overlines denote the
average over all $\ell_{ij}$)%
\be%
\overline{\ell_{\rm OPT}}\sim\sum_{i=0}^d \beta_i
V_{d_i}+\overline{\ell}_{\rm fin} \label{lopt}
\ee%
asymptotically as the size of $\Lambda\to\infty$, keeping the
shape fixed \cite{limit}. Here $\beta_i$ are non-universal
constants (the values of which will change if the $\ell_{ij}$s are
correlated, or for the Euclidean problem), and $V_{d_i}$ are
$d_i=d-i$-dimensional volumes of $\Lambda$ and its boundary. That
is, $V_{d}=|V|$ is the $d$-dimensional volume of $\Lambda$,
$V_{d-1}$ is the $d-1$-dimensional ``area'' of the boundary,
$V_{d-2}$ is the $d-2$-dimensional ``length'' of the edges of the
boundary, \ldots, down to $V_0$, the number of zero-dimensional
corners of $\Lambda$. $\beta_0=\beta$ has been extensively studied
(see e.g.\ Ref.\ \cite{steele} for a review), while bounds on
$\beta_1$ have been established in $d=2$ (Ref.\ \cite{jaillet} for
the Euclidean case). The most interesting part is the subsequent
terms $\overline{\ell}_{\rm fin}$, the leading corrections to the
bulk part of the cost in a finite-size system. These are shape
dependent, and may be difficult to separate from the term $\beta_d
V_0$, since as we will see $\overline{\ell}_{\rm fin}$ can be of
order $1$ for the MST. Here for simplicity we will take $\Lambda$
in the form of a hypercube of side $L$, with periodic boundary
conditions (so all $V_{d_i}$ with $i>0$ are zero). Then we
find as $L\to\infty$ \cite{limit}%
\be%
\overline{\ell}_{\rm fin}\sim -\tilde{\ell}_c+\lambda'L^{\theta}.\ee%
Here $\tilde{\ell}_c$ is the (non-universal) value of the cost of
an edge at the percolation threshold, that is the stage in
Kruskal's greedy algorithm \cite{krus,papst} at which the growing
trees percolate across the system, for $L\to\infty$; in the above
model of $\ell_{ij}$ uniformly distributed in $[0,1]$,
$\tilde{\ell}_c=p_c$, the threshold for bond percolation. Also,
$\lambda'$ is a $d$-dependent non-universal constant. We will
argue that (i) $\theta$ is universal (but depends on $d$), (ii)
$\theta\leq0$ for all $d$,
and (iii) in fact %
\be%
\theta=-1/\nu_{\rm perc},\ee%
where $\nu_{\rm perc}$ is the correlation length exponent for
classical percolation in $d$ dimensions. It is known that
$\nu_{\rm perc}=1$ ($d=1$), $4/3$ ($d=2$), and $\nu_{\rm
perc}=1/2$ for $d\geq d_c$, where $d_c$ is a critical dimension,
$d_c=6$ for percolation; there are approximate values for
$\nu_{\rm perc}$ for other intermediate $d$.

We also consider the effect of a change in boundary conditions. We
can study the mean change in optimum cost produced when a
constraint, that the tree must possess at least $k$ distinct
branches that cross between two ends of the system, for example
between the ends of a cylinder of length $L$ (in one direction)
and width $W$ (in $d-1$ directions), is imposed. We argue that the
mean change in cost per unit length scales as%
\be%
\lim_{L\to\infty}\frac{\overline{\ell_{\rm
OPT}(k)}-\overline{\ell_{\rm
OPT}}}{L}\sim\lambda'_kW^{\theta-1},\label{ellk}\ee%
as $W\to\infty$, for all dimensions $d$, again with
$\theta=-1/\nu_{\rm perc}$.

These finite-size corrections to the mean cost, and its
sensitivity to boundary conditions, are analogous to those for the
ground-state energy of disordered classical systems, such as spin
glasses \cite{theta,bkm}, and the application of such ideas to
optimization was begun in Ref.\ \cite{moore}. It was previously
argued \cite{jrs} for the traveling salesman problem that similar
forms hold in $d=2$ with $\theta$ replaced by $0$ (and with
$L^\theta$ in $\ell_{\rm OPT}$ replaced by a logarithm in some
cases), and should also hold for MSTs. It now appears that the
coefficient $\lambda$ of those terms \cite{jrs} is zero, at least
for MSTs.

The size-dependent terms in $\overline{\ell_{\rm OPT}}$ are
related to the non-zero--temperature behavior of weighted spanning
trees. In this, we give each spanning tree a (Boltzmann-Gibbs)
probability proportional to $e^{-\ell/T}$, where $T$ is the
temperature. The probabilities are normalized by dividing by
the partition function %
\be%
Z=\sum_{{\bf T}\in\cal T}\prod_{\langle ij\rangle\in \bf T}
e^{-\ell_{ij}/T}.
\ee%
In the limit as $T\to 0$, the sum over trees is dominated by those
with the lowest total cost $\ell$. This approach allows methods of
equilibrium statistical mechanics to be applied. We argue that at
a small positive temperature, the entropy per vertex in the limit
as the size of $\Lambda$ tends to infinity, $s$, (essentially the
logarithm of the number of near-optimal spanning trees accessible
at temperature $T$, divided by $|V|$) behaves as%
\be%
s\sim aT^\psi,\ee%
as $T\to0$, where $\psi$ is another universal exponent, most
likely equal to $1$ for MSTs (this has also been discussed in
Ref.\ \cite{ap}). Correspondingly, $\overline{\langle
\varepsilon\rangle}$, the change in the thermal (as well as
$\ell_{ij}$) average cost per vertex relative to the optimum,
is%
\be%
\overline{\langle
\varepsilon\rangle}=\lim_{|V|\to\infty}\frac{\overline{\langle
\ell\rangle}-\overline{\ell_{\rm OPT}}}{|V|}\sim b T^{\psi+1}.\ee
For a large system, $\overline{\langle \varepsilon\rangle}$ is the
thermal and $\ell_{ij}$ average of the notion of ``fractional
relative error'' in optimization theory, within a factor of
$\beta_0$. Inverting these formulas implies that the logarithm of
the typical number of spanning trees with cost within a factor
$1+\varepsilon$ of $\ell_{\rm OPT}$ (where ``typical'' can be made
precise using the Boltzmann-Gibbs probability), divided by $|V|$, is%
\be%
s\sim a'\varepsilon^{\psi/(\psi+1)}\ee %
as $\varepsilon\to0$. Note that these formulas are for the limit
$|V|\to\infty$ before $T\to 0$; the arguments that suggest that
$\psi=1$ also suggest that $s$ and
$\overline{\langle\varepsilon\rangle}$ are dominated by local,
independent excitations, with a density of order $1/T^\psi$, and
so there is a length scale $\xi_T\sim T^{-1/(d\psi)}$ such that
these results hold for system size $L\gg\xi_T$.

In addition to the cost, one may also ask about correlation
properties of the trees, either at $T=0$ (i.e.\ for MSTs), or in
the positive-$T$ generalization. For example, one may consider the
expected number of trees that possess $k$ distinct branches that
cross between two balls separated by distance $r$, as a function
of $r$, and so define correlation exponents (see e.g.\
\cite{ABNW,read}). Another exponent is obtained from the Hausdorff
dimension of the path between two given points on the (same) tree.
These universal exponents serve to distinguish universality
classes. One may ask whether the exponents for the statistics of
the MSTs are the same as for uniform spanning trees. Uniform
spanning trees (USTs) arise if we set all $\ell_{ij}=0$, or put
$T=\infty$, in the positive-temperature weighted spanning trees.
Thus, every spanning tree has equal (``uniform'') Boltzmann-Gibbs
probability. We will argue the following: nonzero temperature is a
relevant perturbation (in the renormalization-group sense), and
leads to a correlation or crossover length $\xi$ ($\xi\gg \xi_T$
for $d>1$), such that for correlation functions over distances
much larger than $\xi$, the behavior of USTs is recovered, even if
$T$ is very small. In an infinite system, this length diverges as%
\be%
\xi\sim c T^{-\nu}\ee%
as $T\to0$. We argue, using results from the extensively-studied
related problem of random (classical) resistor networks (RRNs),
which again is related to percolation, that $\nu=\nu_{\rm
perc}=-1/\theta$. That is, $-\theta$ is the scaling dimension for
the temperature $T$.

These results then imply that if we choose a typical spanning tree
with $\ell$ within about $1+\varepsilon$ of $\ell_{\rm OPT}$, then
its statistical properties on length scales larger than $\xi$ are
those of USTs. The crossover length scale is $\xi\sim c'
\varepsilon^{-\nu/(\psi+1)}$. When $\xi$ is of order the system
size $L$, or on length scales smaller than $\xi$, the correlations
are those of MSTs, which should be different from those of USTs,
at least in high dimensions $d$. Arguments by Newman and Stein
\cite{ns} show that for MSTs, for $d>8$ the MST in any finite
portion of size $W$ of the system breaks up, as
$|\Lambda|\to\infty$, into of order $W^{d-8}$ trees of size of
order $W$, each tree having Hausdorff dimension $8$ (their
arguments also used a relation with percolation). Thus $8$ is a
critical dimension for MSTs, above which the exponents mentioned
above take simple values, related to the Hausdorff dimension $8$
that determines the $k$-crossing exponents, while (by a simple
extension of the arguments of NS) the Hausdorff dimension of the
path between two points becomes $2$, as for a Brownian walk. By
contrast, USTs have similar behavior, but consist of trees of
Hausdorff dimension $4$ for dimension bigger than $4$ \cite{pem}.
However, a relation between the two in low dimensions, in
particular $d=2$, has not been ruled out, and exists, albeit
somewhat trivially, in $d=1$.

It is interesting that the properties of MSTs fall into two parts.
For properties involving the costs, the critical dimension is
argued here to be $d_c=6$. On the other hand, the geometric
correlations of the trees themselves exhibit a critical dimension
of 8. We note that the costs are independent of the tree geometry
in the sense that, given the MST, the costs of the edges used
cannot be recovered (in the lattice models, though this can be
done in the Euclidean case). In the absence of a field theoretic
formulation, analogous to that for equilibrium positive-$T$
critical phenomena, the presence of two distinct critical
dimensions should not seem so surprising.

This paper is structured as follows. Section \ref{sec:mst}
considers the MST problem, and its nonzero temperature
generalization, for large systems. The main results of this
section are the exponent for the crossover length $\xi$,
$\nu=\nu_{\rm perc}$, and the behavior of the entropy and mean
cost (per vertex) at low temperature. In section \ref{sec:fin},
aspects of finite-size systems are considered, first for zero
temperature (MSTs). Using finite-size scaling arguments for
percolation, the two corrections in $\overline{\ell}_{\rm fin}$
are obtained. The change in cost produced by a change in boundary
condition on a long cylinder is considered in section
\ref{sec:boun}. Finally, scaling at both finite size and positive
temperature is considered. Section \ref{sec:oth} considers other
optimization problems, including minimum cost Steiner tree,
traveling salesman, and minimum weighted matching. Some of these
are argued to be in the same universality class as MSTs.


\section{MSTs, RRNs, and percolation}\label{sec:mst}

This section begins with a mapping of the general weighted
spanning tree problem to the calculation of a determinant of a
Laplacian matrix on $G$. The resulting linear-algebra problem is
related to other problems of physical interest, including RRNs.
This problem is then solved as $T\to0$, and related to Kruskal's
greedy algorithm and to a class of corresponding percolation
problems. At nonzero temperature, the connection with RRNs gives
the behavior (as $T\to 0$) of the crossover length $\xi$ to
uniform spanning tree behavior at large length scales. The entropy
and mean extra cost (per vertex) are considered next, and related
to the number of near-optimal spanning trees. Finally some
comments on the mobility edge in the lattice Laplacian are made,
in the strong disorder regime $T\to0$.

\subsection{Mappings between problems}

The partition function $Z$ can be reformulated as a determinant,
by the matrix-tree theorem extended to include
weights $K_{ij} = e^{-\ell_{ij}/T}$ \cite{vlw},%
\be%
Z={\det}' \Delta, \ee%
where $\det'$ denotes the determinant of a matrix, from which any
one row and the corresponding column have been deleted, and
$\Delta=N K N^t$ is defined as follows. $N$ is the incidence
matrix of $G$ viewed as a directed graph by adding an arrow to
every edge in an arbitrary fashion;
then for vertices $i$ and edges $e$,%
\be%
N(i,e)=\left\{\begin{array}{rl} 0&\hbox{ if $i$ is not on $e$,}\\
                                    1&\hbox{ if $i$ is the head of $e$,}\\
                                    {}{-1}&\hbox{ if $i$ is the tail of $e$.}
                                    \end{array}\right.
                                    \ee%
$N^t$ denotes the transpose of $N$, and $K$ is the diagonal
$|E|\times |E|$ matrix with entries
$K(e,e)=K_{ij}=e^{-\ell_{ij}/T}$ for the edge $e=\langle
ij\rangle$.

The matrix $\Delta=N K N^t$ can be regarded as a Laplacian on $G$.
It has a zero mode, the vector $(1,1,\ldots,1)^t$, and is positive
semi-definite (if all $\ell_{ij}/T$ are real), as can be seen by
writing $N'=N K^{1/2}$, and $\Delta = N' N'^t$. The deletion of a
row and column from $\Delta$ before calculating the determinant
removes the zero mode, which would otherwise cause the determinant
to vanish.

Now we suppose, as in the introduction, that the graph $G$ is a
portion $\Lambda$ of a $d$-dimensional lattice, and that the costs
are random variables. Then there are some physical problems that
can be associated with the mathematical system defined by
$\Delta$. For example,
consider the eigenvalue problem for the matrix $\Delta$,%
\be%
\Delta v =\lambda v.%
\ee%
This is similar to the problem of finding the eigenfrequencies
$\pm\sqrt{\lambda}$ for a collection of unit masses connected by
springs with random spring constants $K_{ij}>0$ (but with scalar
rather than vector displacements), or similarly the spectrum of
linearized magnons in a magnet with random exchange constants. The
exact zero mode is associated with the spontaneous breaking of a
symmetry. Such problems have been studied for a long time (see
e.g.\ Refs.\ \cite{dyson,jss,zim} and Ref.\ \cite{gc} contains a
review), although as $T\to0$ the probability distribution for
$K_{ij}$ we consider is particularly broad. The eigenvalue problem
is considered further in the following.

Another problem, which goes back to work by Khirchoff, associated
with this linear system is that of a resistor network. Let
$I=(I_e)$ be the column vector of currents (in the direction of
the arrow) along the edges $e$. In the absence of any external
current sources, the net current into any vertex is zero, that is%
\be%
NI=0.%
\ee%
If potentials $\phi_i$ are associated with each vertex $i$
[forming a column vector $\phi=(\phi_i)$], then Ohm's law states
that
\be%
I=-K N^t \phi, \ee %
where $K_{ij}=(R_{ij})^{-1}$ is the reciprocal of the resistance
(i.e\ the conductance) of the edge $e=\langle ij\rangle$.
Eliminating the currents then gives $\Delta \phi=0$, which of
course is solved by the zero mode, $\phi=$ constant.

If one wishes to find the resistance between any two vertices, by
connecting an external current source across them, then this also
uses the matrix $\Delta$. If a current $J_i$ enters the network at
each vertex $i$, then forming the column vector $J=(J_i)$, we now
have%
\be%
NI=-J \ee so $\Delta \phi=J$ ($\sum_i J_i=0$, otherwise there will
be no solutions). Then%
\be%
\phi= \Delta'^{-1}J%
\ee%
(plus an arbitrary constant), where $\Delta'$ denotes $\Delta$
restricted
to the subspace orthogonal to the zero mode, so that%
\be %
\Delta'^{-1}= {\sum_{n\neq 0}} v_{(n)}v_{(n)}^t/\lambda_n,
\label{propag}\ee%
where $\lambda_n$, $v_{(n)}$, are the eigenvalues and normalized
eigenvectors of $\Delta$, and the zeroth eigenvalue $\lambda_0=0$
is omitted from the sum. From $\phi$, the current flowing along
any edge in the presence of arbitrary sources $J$ can be found.
Then the resistance between vertices $i$
and $j$ can easily be shown to be%
\be%
R_{{\rm (equiv)}ij}=
(\Delta'^{-1})_{ii}+(\Delta'^{-1})_{jj}-2(\Delta'^{-1})_{ij}.
\label{Requiv}\ee

One popular version of the random resistor network problem is that
in which the resistors $R_{ij}$ on the edges are either a constant
$R$, or infinity, with independent probabilities $p$, $1-p$
respectively. This has an obvious connection with percolation
\cite{stah}. In this paper we are instead interested in the case
where $R_{ij}$ has a continuous, but very broad distribution, as
in Ref.\ \cite{ahl}. The specific form in which we are interested,
because of its connection with weighted spanning trees, is
$R_{ij}= e^{\ell_{ij}/T}$, with $\ell_{ij}$ random variables, and
$T$ going to zero (it arises, for example, if $\ell_{ij}$ is the
Euclidean distance between vertices $i$ and $j$ that represent
localized states, $T$ is the localization length, treated as a
constant, and is one aspect considered in Ref.\ \cite{ahl}). This
form also has a less obvious connection with percolation, as we
will see. Our simplest model, in which $\ell_{ij}$ are independent
and uniformly distributed on $[0,1]$, has been studied before
\cite{cgrth,leD,stah,stan}. The distribution of conductances on
the edges is then $P(K_{ij})=T K_{ij}^{-1}$ for
$K_{ij}\in[e^{-1/T},1]$.


\subsection{Solution of eigenvalue problem as $T\to0$}

The next step we will take is to study the eigenvalue problem for
strong disorder, $T$ small, first in the extreme limit as
$T\to\infty$ for a fixed finite graph $G$ with given weights
$\ell_{ij}$. In this limit, the eigenvalues and eigenvectors are
determined by a simple procedure, that is related both to the
greedy (Kruskal \cite{krus}) algorithm which solves the MST
problem \cite{papst}, and to the real-space renormalization group
method for strong disorder that has been applied to quantum
problems (from this point of view, $\Delta$ is the Hamiltonian for
a one-particle hopping problem). Since $\Delta$ contains terms
that vary greatly in magnitude, we may begin by finding the
largest $K_{ij}$, all other terms being negligible compared with
this (since we are interested eventually in the random version
with a continuous distribution, in which with probability one no
two $K_{ij}$ are equal, we neglect the possibility of equal
$K_{ij}$s). Let us relabel the vertices so that those connected by
the largest $K_{ij}$ are $1$ and $2$. At this level of
approximation, the matrix breaks into a $2\times 2$ block, and
$|V|-2$ other $1\times 1$ zero blocks. The $2\times 2$ block has a
normalized eigenvector $(1,-1)^t/\sqrt{2}$ that has eigenvalue
$2K_{12}$, and another eigenvector $(1,1)^t$ with eigenvalue $0$.
Then we find the next strongest $K_{ij}$. This either connects two
vertices (which can be relabeled as $3$, $4$) distinct from 1 and
2, or else it connects either 1 or 2 to a vertex 3 (we may relabel
so that it is $K_{23}$). In the first case, two eigenvectors of
the $3$--$4$ block can be found as for 1 and 2. In the second
case, in the strong disorder ($T\to0$) limit, $K_{12}$ is much
larger than $K_{23}$. We have a situation of degenerate
perturbation theory, in which the eigenvalue $2K_{12}$ has a
negligible correction from $K_{23}$, while the remaining $|V|-1$
orthogonal vectors have zero eigenvalue when $K_{23}$ is
neglected. When $K_{23}$ is included, we derive a reduced
Hamiltonian by projecting the $K_{23}$ terms to the subspace of
zero eigenvalues of the previous step. This contains only one
$2\times 2$ nonzero block, and it turns out that this produces a
nonzero eigenvalue $3K_{23}/2$, with normalized eigenvector
$(1,1,-2,0,\ldots)^t/\sqrt{6}$ in the original basis, as well as a
zero mode $(1,1,1,0,\ldots,)^t/\sqrt{3}$. Hence the subspace of
remaining zero modes has a basis that consists of the latter
vector which involves three vertices that have been connected by
the couplings $K_{12}$ and $K_{23}$, and $|V|-2$ vectors, each for
a single vertex that has not yet been connected. These form the
degenerate subspace within which the next largest $K_{ij}$ must be
considered. Similarly, in the first case, the zero-mode subspace
has a basis that consists of two eigenvectors that involve two
vertices each, and $|V|-4$ that involve one each.

This procedure can be easily iterated. After each step, the space
of remaining zero modes possesses a natural basis with one basis
vector for each of a number of clusters of vertices, which have
been connected by the couplings $K_{ij}$ that were considered at
earlier stages. For each cluster, of say $n$ vertices, the
zero-mode eigenvector is a non-zero constant on those vertices,
and zero elsewhere. The next strongest $K_{ij}$ that has not
already been considered (or ``tested'') must be projected into
this zero-mode subspace. One additional possibility occurs in
general, as the $K_{ij}$ are considered in decreasing order.
Sometimes the next strongest $K_{ij}$ connects two vertices that
already in the same cluster. In this case, the resulting
$1\times1$ block produces an eigenvalue 0 and no change in the
eigenvector. Thus these couplings may be ignored. The interesting
inductive step thus involves a $K_{ij}=K$ that couples two
zero-mode clusters containing, say, $n$ and $m$ vertices
respectively. The projected matrix in the subspace spanned by
these two normalized
eigenvectors takes the form%
\be%
\left(\begin{array}{cc}K/n&-K/\sqrt{nm}\\
                                   -K/\sqrt{nm}&K/m\end{array}\right),
\ee%
and has eigenvalues $(n+m)K/(nm)$, with eigenvector
$(\sqrt{m},-\sqrt{n})^t/\sqrt{n+m}$, and zero, with eigenvector
$(\sqrt{n},\sqrt{m})^t/\sqrt{n+m}$. In the original basis, the
zero-mode eigenvector is again of the form of a constant on the
connected cluster of $n+m$ vertices and zero elsewhere, which
allows the induction to proceed. This procedure can be followed
until $|V|-1$ non-zero eigenvalues have been found, and there is
the one remaining zero mode of $\Delta$ itself, which in the
original basis is $(1,1,\ldots,1)^t/|V|^{1/2}$.

We see that this procedure takes the $K_{ij}$ in sequence,
beginning with the largest (corresponding to the smallest
$\ell_{ij}$), and discarding those that connect vertices that have
already been connected. Hence at each step, the clusters of
vertices formed by the zero modes each take the form of a tree,
connected by the stronger couplings $K_{ij}$ that correspond to
non-zero eigenvalues, but which do not form a cycle. The clusters
form a spanning forest of trees (some trees may contain only a
single vertex and no edges), until the last step at which a single
spanning tree is formed. This procedure of constructing a tree by
adding the lowest-cost edges unless they form a cycle is exactly
Kruskal's greedy algorithm for finding the MST \cite{krus}. To see
that it solves the MST problem, we may construct the partition
function. The determinant $\det'\Delta$ is essentially the product
of the non-zero eigenvalues of $\Delta$. We have shown that this
product is approximately $|V|e^{-\sum_{\langle ij\rangle\in {\bf
T}}\ell_{ij}/T}$, where $\bf T$ is the spanning tree obtained by
the above procedure. The removal of one row and column before
calculating the determinant removes the factor $|V|$. Our approach
has constructed the leading term in the partition function as
$T\to0$, and gives a proof that the greedy algorithm is correct
(there are of course other ways to show that \cite{papst}, without
linear algebra, but the present approach will be useful to us).

\subsection{Connection with percolation}
\label{sec:perc}

It is of interest to study the structure of the eigenvectors of
$\Delta$, especially in a large portion $\Lambda$ of the
$d$-dimensional cubic lattice ($\Lambda$ will be assumed to be a
connected domain with a smooth boundary, such as a cube). First we
establish a connection with percolation. Suppose that the set of
costs $\ell_{ij}$ is given. Then at a step where all edges of cost
$\ell_{ij}<\tilde{\ell}$ have been tested, the clusters formed by
the zero modes can be thought of as (a sample of) bond percolation
clusters (even when a probability distribution on the $\ell_{ij}$
has not been specified). Moreover, if we are only interested in
which vertices are connected in the clusters that represent the
zero modes at a particular step, then it makes no difference to
include the edges that were tested earlier but discarded as they
formed a cycle. Now we will suppose that the $\ell_{ij}$ are
random variables, but not necessarily that the costs for distinct
edges are statistically independent (note that this includes the
Euclidean model, as well as general lattice models). If all edges
with cost $\ell_{ij}\leq \tilde\ell$ are ``occupied'', then we
have a general form of bond percolation, with correlated
bond-occupation probabilities. We will always assume that the
correlations in the $\ell_{ij}$ are short-ranged (falling, say,
exponentially with distance), and translationally-invariant, and
that the cumulative probability for any single $\ell_{ij}$ is
continuous. In percolation, there is a percolation threshold at
$\tilde{\ell}=\tilde{\ell}_c$, such that in the limit $\Lambda\to
{\bf Z}^d$, for $\tilde{\ell}<\tilde{\ell}_c$ any connected
cluster is finite (with probability one), while for
$\tilde{\ell}>\tilde{\ell}_c$ there is a single infinite cluster,
as well as many finite ones (except when $\tilde{\ell}$ reaches
the supremum of the support of the probability density of
$\ell_{ij}$). In the simplest model that we use, which contains
the generic (or universal) behavior of short-range correlated
percolation, the costs $\ell_{ij}$ are statistically independent,
and each is distributed uniformly in $[0,1]$. The corresponding
percolation model is then that in which the bonds (edges of
$\Lambda$) are occupied (independently) with probability
$p=\tilde{\ell}$, and unoccupied with probability $1-p$. The
percolation threshold in this model will be denoted $p_c$. In this
model, in one dimension, $p_c=1$, and in two dimensions $p_c=1/2$
on the square lattice, by duality arguments. In the Euclidean
model of MSTs, each $\ell_{ij}$ is the Euclidean distance between
$i$ and $j$, where the $|V|$ points are (in the simplest Euclidean
model) independently and uniformly distributed over the domain
$\Lambda$ (with density 1). In this model, the corresponding
percolation problem becomes (the Voronoi, or ``lily pad'',  form
of) continuum percolation.

In the simplest, independent-edge, model of bond percolation, the
finite clusters above and below $p_c$ have typical size $\xi_{\rm
perc}$ which diverges at $p\to p_c$ as $\xi_{\rm perc}(p)\sim
|p-p_c|^{-\nu_{\rm perc}}$, where $\nu_{\rm perc}$ is a universal
$d$-dependent exponent. As $p\to p_c$, these typical clusters are
fractals with Hausdorff dimension $D_{\rm perc}$. For $d>6$,
$\nu_{\rm perc}=1/2$ and $D_{\rm perc}=4$; the clusters behave as
branched polymers (trees) with no, or negligibly many, cycles
(even though cycles are not forbidden in percolation). These
properties are also believed to hold, with the same exponents, for
the more general models with short-range correlations of the
$\ell_{ij}$, with $\tilde{\ell}$ ($\tilde{\ell}_c$) in place of
$p$ ($p_c$, respectively), provided that the probability density
for each single $\ell_{ij}$ is smooth at $\tilde{\ell}_c$.
$\tilde{\ell}_c$ is non-universal, that is it depends on the
details of the probability distribution. In the following, results
will be given in terms of the simplest model, but hold equally for
the other models.

The relation we have described of the growing trees in Kruskal's
algorithm to percolation is similar to that
\cite{ns,cieplak1,bara,cieplak2,dd} between Prim's algorithm
\cite{prim,papst} (which for a given finite sample ultimately
produces the same MST) and invasion percolation \cite{inv}.
Invasion and ordinary percolation (at the percolation threshold)
are believed to be in the same universality class.

The eigenvectors with non-zero eigenvalues are always a
combination of two clusters from the preceding step in the
algorithm, that are connected by the next-strongest coupling
$K_{ij}$, with amplitudes that are constant on each of the two
clusters. More precisely, the amplitudes are %
\be%
\frac{1}{\sqrt{n+m}}\sqrt{\frac{m}{n}}\ee%
for each vertex on the cluster of $n$ vertices, and minus the same
but with $n$ and $m$ interchanged on the cluster of $m$ vertices.
Hence, for $\ell_{ij}<p_c$, where both clusters typically have
size of order $\xi_{\rm perc}$ (evaluated at $p=\ell_{ij}$), the
eigenvector is localized on a length scale also of order $\xi_{\rm
perc}$. For $\ell_{ij}>p_c$, there is an infinite cluster, i.e.\
one that occupies a finite fraction of the vertices as $\Lambda\to
{\bf Z}^d$. In this case, by letting $n\to\infty$, we find that
the normalized eigenfunction is concentrated on the finite cluster
of $m$ vertices, and so is also localized, with localization
length diverging as $\xi_{\rm perc}$ as $p\to p_c$. Thus, with the
exception of the zero mode, in the strong disorder limit all
eigenvectors of $\Delta$ are localized, except at $p\to p_c$ where
the localization length diverges. The mean localization length
presumably increases monotonically as the $\ell_{ij}$
corresponding to the eigenvalue increase to $p_c$, then for
$\ell_{ij}>p_c$ decreases monotonically as $\ell_{ij}\to 1$.


\subsection{Effective resistance in the strong disorder limit}

We now apply the preceding results to the effective resistance
between any two vertices, $R_{({\rm equiv})ij}$, using eqns.\
(\ref{Requiv}), (\ref{propag}), in the strong disorder ($T\to 0$)
limit.

Each eigenvector has the structure described in the previous
Section, with constant amplitude on two clusters of sizes $n$, $m$
connected by the next strongest coupling, $K$ say, and zero
elsewhere, and can only contribute to $R_{({\rm equiv})ij}$ if at
least one of $i$, $j$ lies on one of the clusters. Suppose there
is nonzero amplitude at both $i$ and $j$. If both are in the same
cluster, then the contributions to $R_{({\rm equiv})ij}$ cancel.
If they are on opposite clusters, the contribution to $R_{({\rm
equiv})ij}$ is%
\be%
\frac{1}{m+n}\left(\frac{m}{n}+\frac{n}{m}+2\right)
\frac{nm}{(n+m)K}\ee which simplifies to $1/K$. Finally, if one of
$i$, $j$, say $i$, is on a cluster (say, the one of $n$ vertices)
but the other $j$ is not, then the contribution is
$m^2/[(n+m)^2K]$.

In the procedure that generates the eigenvectors, the sizes of the
clusters are monotonically increasing. For given $i$ and $j$, the
situation that one of $i$, $j$ is on one cluster, the other on the
other occurs only once, at the stage where those two clusters get
connected, so there is only a single contribution of the form
$1/K$. The situation where only one of $i$ and $j$ is in a cluster
occurs at {\em larger} values of the couplings than this $K$. For
smaller couplings than $K$, both vertices are both in the same
cluster, or neither is on a cluster. Then as $T\to0$, this single
term $1/K$ dominates the equivalent resistance. This is consistent
with the picture that in the strong disorder limit, the current
from $i$ to $j$ is carried along a single non-self-intersecting
path of edges, such that the sum of resistances along the path is
minimized. However, the total resistance of a path is dominated by
the largest resistance on the path, and this is exactly the
resistance $1/K$.

We see that the current must pass through the edge of resistance
$1/K$ that we have singled out, in a particular direction that is
also determined (this could be verified also by calculating the
current on any edge, using formulas from the previous section).
Then the current injected at $i$ must pass along the edges to the
correct end of this edge. In the strong disorder limit, we may use
the above arguments again to find the resistance between these
vertices, which is again dominated by a single resistor of
resistance $<K^{-1}$. This construction can be repeated until the
complete path of lowest resistance from $i$ to $j$ has been found.
Each resistor on the path is one of those that corresponds to a
non-zero eigenvalue of $\Delta$, and so lies on the MST. It
follows that {\em in the strong disorder limit, the path of least
resistance between any two vertices lies along the MST}. In other
words, the MST is the solution to the following problem (the
all-pairs minimax path problem) \cite{hu}: given a ``resistance''
on each edge of a connected graph $G$, for each pair of vertices
$i$, $j$, find the path from $i$ to $j$ that minimizes the value
of the largest resistance on the path, and take the union of these
paths over all pairs of vertices $i$, $j$.


\subsection{Effect of small nonzero temperature}

Now we turn to the behavior at a small non-zero temperature $T$,
which means a finite strength of disorder; here, we present
arguments using only percolation theory, leaving the behavior of
the eigenvalue problem for a later section.

RRNs in $d$ dimensions with resistances of the form
$R_{ij}=e^{\ell_{ij}/T}$ for $\langle ij \rangle$ an edge
connecting nearest neighbors, $R_{ij}=\infty$ otherwise, have been
considered in several earlier works \cite{ahl,cgrth,leD,stan}. As
the distribution of resistances is very broad for $T$ small, the
following picture of the network emerges. If we consider the
clusters that are connected by resistors with
$\ell_{ij}<\tilde{\ell}$, then for $\tilde{\ell}<\tilde{\ell}_c$
these do not percolate. They consist of low resistances, which can
be considered to be essentially zero (like superconducting links).
Resistors with $\tilde{\ell}_c-T<\ell_{ij}<\tilde{\ell}_c+T$ (the
exact coefficient of $T$ in these bounds is not precisely defined,
but is order 1, and is set to 1 for illustration) are all of a
similar magnitude, and connect the superconducting clusters into a
network that spans a positive fraction of the system. Finally, the
resistors with $\ell_{ij}>p_c+T$ connect other clusters to this
network, but these clusters are shorted out by the lower resistors
and do not contribute to conduction on large scales. On large
scales, the resistance or conductance of the system is that of an
effectively uniform medium described by a conductivity $\sigma$
(note that the conductance [the reciprocal of the resistance] of a
cube of size
$L$ is $\Sigma=\sigma L^{d-2}$), with %
\be%
\sigma \propto e^{-\tilde{\ell}_c/T}
\times\left\{\begin{array}{ll}
           T^{(d-2)\nu_{\rm perc}},&\hbox{ $d\leq 6$,}\\
                                    T^2,&\hbox{ $d\geq 6$.}
                                    \end{array}\right.\label{corrlen}\ee%
This arises as follows: there is a conductance of around
$e^{-\tilde{\ell}_c/T}$ for each ``critical'' edge \cite{ahl}, and
negligible resistance for the clusters connected by these edges.
Then for dimensional reasons, the density of critical edges that
connects clusters contributes a factor of length to the $d-2$
power, and this length must be the size of the clusters used,
which is $\xi_{\rm perc}(p=\tilde{\ell}_c-T)\propto T^{-\nu_{\rm
perc}}$. For $d>6$, there is an additional power $\xi_{\rm
perc}^{d-6}$ which is the number of distinct connected percolation
clusters in a window of size $\xi_{\rm perc}$ at criticality
\cite{stah} (this number is of order one for $d\leq 6$---this is
the breakdown of hyperscaling relations for $d>6$, expressed in
terms of the geometry of the clusters \cite{stah,aizen}); these
distinct conducting channels add since they are in parallel. The
possibility of an additional power of $T$ (as would occur in some
different models of RRNs \cite{stah}) was investigated, and bounds
on its exponent were found \cite{cgrth}. Le~Doussal \cite{leD}
argued that the power of $T$ in $\sigma$ is exactly as given in
eq.\ (\ref{corrlen}). It should be noted that in these earlier
works the length scale above which the effective medium, with
negligible fluctuations in conductivity, applies is $\xi_{\rm
perc}\propto T^{-\nu_{\rm perc}}$. This length scale has also been
identified in a recent work that examined finite-size scaling
properties of the RRN \cite{stan}. This length scale is an
important result for
weighted spanning trees (i.e.\ MSTs at positive $T$) as well: %
\be%
\xi\propto T^{-\nu},%
\ee%
with $\nu=\nu_{\rm perc}$.

\subsection{Cost and entropy at positive temperature}\label{posT}

In this section, we address the positive-temperature properties of
weighted spanning trees directly, that is in terms of trees, not
resistor networks.

The most elementary excitation of a spanning tree is to move an
edge. By this we mean that an edge on the tree is removed, thus
cutting the tree into two parts, which are then reconnected by
adding a different edge (not on the initial tree). The change in
cost is simply the difference of the costs of the two edges
involved. All spanning trees can be reached from the MST by
successive operations of this type \cite{papst}. Starting from any
spanning tree, then because our models assume a continuous
distribution of costs for the edges, with probability one either
it is the unique MST, or it is possible to move one edge such that
the total cost {\em decreases} \cite{papst1}. Hence there are no
true ``metastable states'' (i.e.\ local, but not global, minima
with respect to moving a single edge) in the MST problem, at least
not on a finite graph as assumed in these arguments.

At low temperatures $T$, there will be thermal excitation of
single-edge moves, which can occur independently. Consider the
following situation. In the greedy algorithm, suppose that edge
$\langle ij\rangle$ is added to the MST when
$\ell_{ij}=\tilde{\ell}$. Suppose further that, before this edge
is added, the trees (clusters) already grown are such that
$\langle ij\rangle$ and one other distinct edge $\langle kl
\rangle$ would form a cycle if both were added. Then adding
$\langle ij\rangle$ to the tree prevents the subsequent addition
of $\langle kl \rangle$. But if $\langle kl \rangle$ is added
instead of $\langle ij\rangle$, then this connects the same two
clusters, and hence does not affect which edges can be added at
subsequent stages of the greedy algorithm, that is at
$\tilde{\ell}>\ell_{ij}$. As such pairs of edges will be found at
all stages of the greedy algorithm, there will be many such pairs
of edges in the MST, each of which may be excited (one edge
replaced by its partner) independently of the states of the other
pairs.

If we consider only these pairs, then a simple picture of
``two-level systems'' (TLSs) \cite{ahv} emerges, that should be
useful at low $T$: other than the MST, the spanning trees that
contribute to the partition function differ from the MST only by
having one or more of the edge-pairs excited, and these can be
excited independently. The partition function within this picture
can be calculated easily, if the excitation costs
$\ell_{kl}-\ell_{ij}$ are given. One needs some information about
the probability distribution of these excitation costs. Let us
consider only values of $\ell_{ij}$ (as before, this is the lower
of the costs for each pair) that are bounded away from the
critical value $\tilde{\ell}_c$. Then the sizes of the clusters
connected by $\ell_{ij}$ at that stage of the greedy algorithm are
of order $\xi_{\rm perc}(\ell_{ij})$ [for cases where
$\ell_{ij}>\tilde{\ell}_c$, we mean the size of the finite
cluster(s) involved], which is bounded. When $T$ is small, the
pairs in which we are interested have $\ell_{kl}-\ell_{ij}$ of
order $T$ or less, and occur with density tending to zero with
$T$; hence the mean spacing between them is much larger than their
size in this limit. It is reasonable to imagine that their
excitation costs are statistically independent, and that the
probability density for the excitation cost of each approaches a
constant as $\ell_{kl}-\ell_{ij}\to0$ (the constant might depend
on $\ell_{ij}$, but this is not important). For example, one can
estimate this probability density, and check statistical
independence of distinct TLSs, for the case when the cycle
involved is an elementary square of side 1. We introduce the
standard (canonical ensemble) statistical mechanics definitions of
the free energy $F=-T\ln Z$, entropy $S=-\partial F/\partial T$,
and internal energy (or cost) $E=F+TS=T^2\partial \ln Z/\partial
T$. The entropy can be thought of as the logarithm of the number
of trees with cost less than the corresponding value of $E$ (this
microcanonical-ensemble definition will agree with the canonical
definition in the limit $|V|\to\infty$ with $S$ and $E \propto
|V|$, and with $T$ fixed, as used here). It follows from the TLS
model that at temperature $T$, the entropy per vertex,
$s=\lim_{|V|\to\infty}S/|V|$, behaves as %
\be s\propto T^\psi\ee%
as $T\to0$, while the thermal average excitation cost per vertex
$\langle \varepsilon \rangle=\lim_{|V|\to\infty}(E-\ell_{\rm
OPT})/|V|$ behaves
as%
\be%
\langle \varepsilon\rangle \propto T^{\psi+1}\ee%
in the same limit, where $\psi=1$. Since we have included only a
subset of the possible excitations, these statements should be
taken as a lower bound on $s$, so that $\psi\leq 1$. This notion
of TLS is generic for many disordered systems \cite{ahv}, and the
behavior $s\propto T$ is typical for these applications. (A
similar picture of TLSs for MSTs was also used in Ref.\ \cite{ap}
to obtain the behavior of the cost of the minimum spanning tree
that differs from the global MST by a given fraction of edges.)
Note that in these statements we did not need to explicitly
perform the disorder average, as the thermodynamic $|V|\to\infty$
limit of these quantities self-averages.

In this argument, we used only TLSs that demonstrably were
completely independent as excitations. There could of course be
other low-energy TLSs, possibly involving moving more than one
edge, that can only be excited conditionally on the states of
other edges. But in general, by a TLS we will mean a {\em compact}
(localized) excitation. We note that the above arguments do not
apply in the one-dimensional case, which however can be solved
directly. For a system of $L$ vertices with a periodic boundary
condition, the entropy and mean excitation cost are of order $\ln
LT$ and $T$ (not $\propto L^d$, unlike the $d>1$ cases),
respectively, as $L\to\infty$ with $T$ fixed; they can be
calculated exactly for the simplest model of independent edges
each distributed uniformly in $[0,1]$.

So far we were careful to move edges that were not close in cost
to the percolation threshold. Now we examine these in detail,
using the simplest model for which the corresponding percolation
problem is uncorrelated bond percolation model. The idea is
similar to that used in the RRN point of view in the previous
section. If we run the greedy algorithm until all edges with cost
less than $p_c$ minus of order $T$ have been tested, then we
obtain a set of clusters of size less than about $\xi_{\rm
perc}(p=p_c-T)$. If we add all edges of cost between this limit
($p_c-T$) and $p_c$ plus of order $T$, then we obtain a giant
cluster that contains a nonzero fraction of the vertices as
$|V|\to\infty$. We are interested in the subset of these edges
that connect distinct components (which can be viewed either as
clusters or as trees) of the spanning forest for
$\tilde{\ell}=p_c-T$; we call these critical edges. Clearly not
all of these critical edges can be on the MST. But for the
positive-temperature weighted spanning tree problem, the many
different ways of adding a subset of the critical edges so as to
obtain only trees have similar Boltzmann-Gibbs weight. We can
construct a reduced graph that has the critical edges as its edge
set, and the connected components for $\tilde{\ell}=p_c-T$ as its
vertices. We will assume that the reduced graph is connected. Then
if we sum over all spanning trees of this reduced graph with the
corresponding Boltzmann-Gibbs weights, then as the differences in
cost are only of order $T$ when any one edge is moved, this
problem is approximately a uniform spanning tree problem. It is
essentially counting all the spanning trees. As in the TLS
argument, the choice of a spanning tree on the reduced graph does
not affect the remaining edges to be added of still higher cost,
which complete a spanning tree of $G$, because the spanning trees
of the reduced graph all connect the same vertices of $G$.

The connectivity properties, such as the probability that $k$
distinct branches of the tree cross between two chosen balls (as
discussed in the introduction), and corresponding scaling
dimensions and Hausdorff dimensions are unaffected by TLSs of size
smaller than the scale on which these correlations are studied.
But on scales larger than $\xi_{\rm perc}(p_c\pm T)$, the argument
here, which is essentially a coarse-graining or renormalization
group argument, suggests that the connectivity properties become
those of uniform spanning trees (USTs). In the UST problem, which
corresponds to the $T\to\infty$ limit of the weighted spanning
trees, disorder (randomness) in the costs $\ell_{ij}$ can be shown
to be irrelevant, that is it has no effect on the large-scale
universal properties. As we have seen that temperature is a
relevant perturbation of the zero-temperature (MST) limit, it
makes sense that the crossover is to USTs at large length scales.
This is consistent with the arguments of the previous section, in
which the conductivity at large scales becomes essentially
non-random, because we can identify the non-random $T\geq0$
(uniform) spanning tree problem with a resistor network with a
constant resistor on each edge of the lattice. We see again that
the crossover length scale diverges as $\xi\propto T^{-\nu}$ as
$T\to0$, with $\nu=\nu_{\rm perc}$. As noted in the Introduction,
by using the above results for the cost and entropy, this can be
interpreted as saying that for a typical spanning tree that has
cost within $1+\varepsilon$ of $\ell_{\rm OPT}$, the length scale
is $\xi\propto \varepsilon^{-\nu/(\psi+1)}$ at $|V|\to\infty$, for
$\varepsilon\to0$.

We should emphasize that saying that temperature is a relevant
perturbation of the zero-temperature MST fixed point does not, in
our view, entirely rule out the possible equivalence of the
universality classes of statistical connectivity properties in the
MST at $T=0$ and USTs. That is because the averages are different
in the two cases. For the MST, we mean the average of a quantity
over the random costs with respect to which the optimum must be
found. For the UST, there is a nonzero (or even infinite)
temperature. Theoretically, it still appears possible that the
universality classes for geometric or connectivity properties are
the same, in sufficiently low dimensions (indeed, in $d=1$ the
resulting probability distributions on trees are the same, though
the connectivity properties are trivial). For $d=2$, this would
imply conformal invariance of the MST. However, the universality
classes for $d=2$ have been compared numerically by looking at
certain exponents \cite{dd,middle}, and while the early results
may not have ruled out their equality, recent numerical evidence
\cite{wilson} seems also to be {\em against} these universality
classes being the same, and against the conformal invariance of
the $d=2$ MST.

The reduced-graph (or coarse-graining) idea can be used to
estimate the contribution to the entropy of the network of
critical edges. On large length scales, the reduced graph behaves
as a finite-dimensional system. Hence, the entropy of the uniform
spanning trees formed using the critical edges only should be of
order the number of vertices of the reduced graph. For $d\leq 6$,
there is of order one connected percolation cluster per
correlation volume $\xi_{\rm perc}(p_c-T)^d$, and hence the
contribution to the entropy per vertex is $\xi_{\rm perc}(p_c\pm
T)^{-d}\propto T^{d\nu}$ for $d>1$. For $d>6$, there are of order
$\xi_{\rm perc}^{d-6}$ connected clusters per correlation volume
\cite{stah}. Hence we expect that the contribution to the entropy
per vertex is $\xi_{\rm perc}(p_c\pm T)^{-6}\propto T^3$ for
$d>6$. In either case, the result is smaller than $T$ as $T\to0$
when $d>1$. For
$2\leq d<6$, we predict then that the entropy per vertex has the form%
\be%
s\sim aT+a_1T^2+a_2 T^{d\nu}+\ldots\ee%
as $T\to0$, where $a$, $a_1$, $a_2$ are non-universal
coefficients. This form can be viewed as an ``analytic part'', in
integer powers of $T$, which we have continued to order $T^2$
(because $d\nu>2$ for $d\geq 2$), plus a non-analytic or singular
part $T^{d\nu}$. (Such a form is familiar from ordinary critical
phenomena at nonzero temperature.) The free energy per vertex,
$f=\lim_{|V|\to\infty}F/|V|$ divided by temperature has a similar
expansion, as does the internal energy per vertex over
temperature, only the coefficients $a$, $a_i$ being changed in
obvious ways in each case. Thus, the earlier arguments that the
leading term $T^\psi$ in $s$ in fact has $\psi=1$ is an argument
that the leading effects are localized excitations that contribute
to the analytic part. A power $\psi<1$ would be viewed as a
non-analytic part, and would presumably indicate that the leading
contribution is from large-scale collective excitations. The
singular part $T^{d\nu}$ for $d<d_c$ is of the form expected when
hyperscaling applies in critical phenomena, except that here it
applies to $F/T$ instead of to $F$. That is because we dealing
with a fixed point (or critical point) at zero temperature, and
the natural quantity that scales is $F/T$, which controls the
probabilities of different configurations (whereas at a transition
at nonzero $T=T_c$, one can expand $F/T_c$ in powers of $T-T_c$).
Hence we expect on general grounds that these expansions are of
the correct form. For $d\geq d_c=6$, the singular part takes the
form $T^3$ which apparently we cannot distinguish unambiguously
from the analytic part. This difference from ordinary critical
phenomena occurs because only $T\geq 0$ is available.

\subsection{Implications for the eigenvalue problem at $T\neq 0$}

In this section, we apply the results obtained in previous
sections from RRNs and from percolation to the eigenvalue problem
for the matrix $\Delta$, in the regime of strong but finite
disorder, $T$ non-zero and small. The results of this section are
not used elsewhere in this paper.

As we saw, when $T\to0$ in a large system, de-localized
eigenvectors (other than the zero mode) occur only at the
percolation threshold $p_c$. On the other hand, when $T$ is
non-zero there is a well-defined probability density for the
$K_{ij}$s. One then expects de-localized (in fact, extended)
eigenvectors to occur at sufficiently low eigenvalues if $d>2$,
while for $d=2$, the localization length diverges as the
eigenvalue $\lambda \to 0$ \cite{jss}. We also expect that for
$d>2$, in the strong disorder limit as $T\to0$, the fraction of
extended eigenvectors tends to zero. One would like to understand
how these two descriptions of the spectrum are connected in the
limit. We will present a partial answer to this question.

When $T$ is small and non-zero, the method used for $T\to0$ breaks
down when the assumption that $K_{ij}$s for distinct edges are
very different breaks down. A typical way for this is to happen is
provided by the configurations that gave the TLS in the previous
section. When $\ell_{ij}$ and $\ell_{kl}$ connect the same two
clusters (zero modes of couplings stronger than either of these),
and are within $T$ of each other, then both must be included in
the reduced $2\times 2$ block, and the eigenvectors and non-zero
eigenvalue they produce are modified, though the eigenvector is
still localized. This does not affect later eigenvectors, and in
the partition function produces the thermal effects we have
described using the TLS picture.

It is very plausible that the extended eigenvectors for small $T$
are produced by the critical edges only, that is those with
$\ell_{ij}$ within $T$ of $\tilde{\ell}_c$ that connect clusters
of size of order $\xi$. This is connected with the crossover to
the UST behavior at large length scales, and to the effectively
uniform conducting medium in the RRN point of view. Hence, one
expects that using these clusters, on length scales larger than
$\xi$, the Laplacian $\Delta$ can be represented by $\Delta_{\rm
eff}=-\sigma \nabla^2$. Then the density of eigenvalues $\lambda$
(per unit volume and per unit $\lambda$) is predicted to be
$\propto \sigma^{-d/2}\lambda^{(d-2)/2}$ as $\lambda\to0$. This
appears to be consistent with other approaches for the $d=1$ case,
which is essentially soluble \cite{dyson,zim} (and the value of
$\sigma$ can also be easily verified for this case \cite{leD}).

Next, there is the question of the behavior of the mobility edge
(the value of $\lambda$ above which, in a large system,
eigenvectors are localized), or alternatively the fraction of
eigenvectors that are extended. We will not enter into a full
study of the spectrum here, but only make a crude estimate, which
may capture the correct asymptotic behavior. Using $\Delta_{\rm
eff}$, we expect that the number of states (per unit volume) with
eigenvalue less than $\lambda$ scales as $\propto
\sigma^{-d/2}\lambda^{d/2}$ as $\lambda\to0$. $\Delta_{\rm eff}$
is valid only for scales $>\xi$, so this can hold only until the
number of states it predicts reaches $\xi^{-d}$. This gives a
``critical'' value for $\lambda$,%
\be%
\lambda_c\propto
e^{-\tilde{\ell}_c/T}\times\left\{\begin{array}{ll}
           T^{d\nu},&\hbox{ $d\leq 6$,}\\
           T^3,&\hbox{ $d\geq 6$,}
                                    \end{array}\right.\ee
as $T\to0$. This value is our prediction for the mobility edge for
$d>2$, though it is possible that the correct value is larger for
$d>6$ because the number of clusters of size $\xi$ that can be
used to construct the extended states is of order $\xi^{-6}$ per
unit volume, not $\xi^{-d}$. The exponential dependence,
$e^{-\tilde{\ell}_c/T}$, agrees with the fact that in the $T=0$
limit, delocalization occurs only at
$\tilde{\ell}=\tilde{\ell}_c$, so only the sub-exponential
dependence on $T$ can be in question.

The fate at $T\neq0$ of the eigenvalues of the $T\to0$ limit at
$\ell_{ij}>\tilde{\ell}_c$ is a puzzle. They should remain
localized, but their density of states appears to overlap that of
the extended eigenvectors. We cannot resolve this here, and so our
description of the spectrum for $d>2$ and for small $T\neq0$ must
remain somewhat tentative.

\section{Finite-size and boundary-condition effects on the total
cost}

In this section we consider the effect of finite system size on
the optimum cost of the spanning tree, and of changing the
boundary conditions (imposing additional constraints) on this
minimum cost. The arguments are largely independent of those in
the last section, except that the relation to percolation again
appears. The results take the form of a term in the subleading (in
inverse powers of system size, $L$ say) behavior of the cost that
features an exponent $\theta$, which is again related to
percolation, $\theta=-1/\nu_{\rm perc}$. Finally, we obtain a
scaling form for the free energy, which exhibits the crossover
between the zero temperature cost and the infinite-size limit at
fixed positive temperature, which is related to the results of the
previous section.

\subsection{Finite-size scaling of the mean cost}\label{sec:fin}

The relation of MSTs to percolation was explained in Section
\ref{sec:perc}. In the most general case, when all edges of cost
less than $\tilde{\ell}$ are occupied, we have a subgraph of $G$
which consists of one or more connected components, called
clusters (there may be clusters consisting of a single vertex and
no edges). This number will be denoted ${\cal N}(\tilde{\ell}|G)$,
and depends implicitly on the set of edge costs $\ell_{ij}$. For
$\tilde{\ell}<\min_{\langle ij\rangle}\ell_{ij}$, ${\cal
N}(\tilde{\ell}|G)=|V|$, and for $\tilde{\ell}>\max_{\langle
ij\rangle}\ell_{ij}$, ${\cal N}(\tilde{\ell}|G)=1$. Between these
limits, ${\cal N}(\tilde{\ell}|G)$ obviously has a sequence of
downward steps of unit magnitude. The MST for the same graph $G$
with the same set of costs consists of those edges which, as
$\tilde{\ell}$ is increased from its lower to its upper limit,
decrease the number of connected clusters by 1. Then we have the
general formula for the optimum cost ({\em without}
averaging):%
\be%
\ell_{\rm OPT}=-\int_{-\infty}^\infty d\tilde{\ell}\,
\tilde{\ell}\, \frac{\partial{\cal N}(\tilde{\ell}|G)}{\partial
\tilde{\ell}}.\ee %
It follows that
the mean cost of the MST is {\em exactly}%
\be%
\overline{\ell_{\rm OPT}}=-\int_{-\infty}^\infty d\tilde{\ell}\,
\tilde{\ell} \frac{\partial\overline{\cal
N}(\tilde{\ell}|G)}{\partial
\tilde{\ell}},\ee %
where $\overline{\cal N}(\tilde{\ell}|G)$ is the mean number of
connected clusters in the corresponding percolation problem. (This
idea is certainly known to probabilists, and is contained in
Frieze's \cite{frieze} exact calculation of $\overline{\ell_{\rm
OPT}}$ as $|V|\to\infty$ for the case of the complete graph [i.e.
one edge $\langle ij\rangle$ for every pair $i$, $j$ of vertices]
with independent costs for the edges.) For the simplest model, in
which the costs are independent and uniformly distributed in
$[0,1]$, this reduces to
\be%
\overline{\ell_{\rm OPT}}=-\int_0^1 dp\, p
\frac{\partial\overline{\cal N}(p|G)}{\partial p},\ee%
which we use hereafter. For the complete graph, the result as
$|V|\to\infty$ is \cite{frieze} $\overline{\ell_{\rm
OPT}}=\zeta(3)$, where $\zeta$ is the Riemann zeta function. In
this paper, we specialize to graphs $G$ that are a portion
$\Lambda$ of a cubic lattice in $d$ dimensions, and we will
further assume here that $\Lambda$ is a cube of side $L$ (parallel
to the lattice axes), with periodic boundary conditions. For this
system, we write the mean number of percolation clusters as
$\overline{\cal N}(p,L)$. Again, the results found below also
apply to the more general models as delimited in the previous
section. The following arguments could be extended further to
study the boundary terms in eq.\ (\ref{lopt}), or further
finite-size corrections.

The function $\overline{\cal N}(p,L)/L^d$ should have a
well-defined monotonically-decreasing limit:%
\be%
Y(p)=\lim_{L\to\infty}\overline{\cal N}(p,L)/L^d,\ee%
where the limit is taken with $p$ fixed. Thus%
\be%
\beta=-\int_0^1dp\,p\frac{\partial Y(p)}{\partial p}.\ee%
The expected fraction of edges of cost between $p$ and $p+dp$ that
lie on the MST as $L\to\infty$ is %
\be%
-\frac{1}{d}\frac{\partial Y(p)}{\partial
p}dp\ee%
for the (hyper-)cubic lattice; this function has been calculated
and plotted in Ref.\ \cite{dd} for some lattices in dimensions
$d=2$ and $3$ (though without making this connection with
percolation, and the singular contributions we discuss below are
not visible). There is a simple but important relation involving
$Y$, which originates from the facts $\overline{\cal N}(1,L)=1$,
$Y(1)=0$. It can be written as:%
\be%
-\int_0^1 dp\, \left(\frac{\partial\overline{\cal
N}(p,L)}{\partial p}-L^d \frac{\partial Y(p)}{\partial
p}\right)=-1,\label{ident}\ee%
and will be used below.

We may now substitute these forms to obtain a result for
$\overline{\ell_{\rm OPT}}$:%
\ba%
\overline{\ell_{\rm OPT}}&=&\beta L^d-\int_0^1
\!\!dp\,p\left(\frac{\partial\overline{\cal N}(p,L)}{\partial
p}-L^d \frac{\partial Y(p)}{\partial p}\right)\nonumber\\
&=&\beta L^d-p_c \nonumber\\
&&{}-\int_0^1 \!\!dp\,(p-p_c)\left(\frac{\partial\overline{\cal
N}(p,L)}{\partial
p}-L^d \frac{\partial Y(p)}{\partial p}\right),\label{formul}\ea%
using eq.\ (\ref{ident}). Notice that in more general models, the
term $-p_c$ is replaced by the value $-\tilde{\ell}_c$ of the cost
at the percolation threshold, as claimed in the introduction.
Next, we present arguments that the remaining integral goes to
zero as $L\to\infty$, and find its magnitude.

In percolation, $\overline{\cal N}(p|G)$ plays the role of the
free energy of a statistical mechanics problem \cite{stah} (this
can be made precise by using the relation of percolation to the
$Q\to1$ limit of the $Q$-state Potts model on the arbitrary graph
$G$)). In the case of a lattice in dimension $d$, $Y(p)$ has a
singular (nonanalytic) behavior at $p=p_c$ ($p_c$ is the
percolation threshold of the {\em infinite} system), which for
$d\geq 2$ has the form
\cite{limit} %
\ba%
Y(p)&\sim
&Y(p_c)+(p-p_c)Y'(p_c)+\frac{1}{2}(p-p_c)^2Y''(p_c)\nonumber\\&&{}+C_\pm
|p-p_c|^{2-\alpha}+\ldots\ea%
as $p\to p_c$. Here $\alpha$ is another universal exponent, $C_-$,
$C_+$ are non-universal $d$-dependent constants for the cases
$p<p_c$, $p>p_c$, respectively, and the leading terms on the right
hand side vanish more slowly than $|p-p_c|^{2-\alpha}$. For
$d<d_c=6$, $2-\alpha=d\nu_{\rm perc}$ (and apparently varies
monotonically), while $2-\alpha=3$ for $d\geq6$. As
$2<2-\alpha\leq 3$ when $d\geq 2$, the non-analytic part of $Y$
does not necessarily contradict the monotonic decrease of $Y(p)$
with increasing $p$. We will define
\be%
Y(p)_{\rm sing}= C_\pm |p-p_c|^{2-\alpha},\ee %
for all $p$, so as to match the non-analytic behavior; $Y(p)_{\rm
sing}$ will be used only in the vicinity of $p_c$. For $d=1$,
$p_c=1$, $\nu_{\rm perc}=1$, $Y(p)=1-p$, and the singular piece
cannot be separated from the background, though $Y(p)$ does obey
the expected linear form as $p\to p_c$ ($Y$ must be positive, so
cannot be smoothly continued to $p>1$; this can perhaps be viewed
as a non-analyticity).

The idea for completing an estimate of the final integral in eq.\
(\ref{formul}) is that the difference of derivatives in the
integrand, which must obviously be smaller than $L^d$ as
$L\to\infty$, is in fact much smaller, and concentrated at
$p=p_c$. At finite $L$, $L^d\partial Y/\partial p$, which has a
nonanalyticity at $p_c$, is replaced by $\partial \overline{\cal
N}(p,L)/\partial p$, which is analytic in $p$ for all $p$ (in
fact, it is a polynomial in $p$). The derivative of the number of
clusters is sensitive to the finite size of the system only
through correlation effects. Consequently, sufficiently far from
$p_c$ that $L\gg \xi_{\rm perc}\propto |p-p_c|^{-\nu_{\rm perc}}$
as $p\to p_c$, the difference between the two functions is of
order $e^{-c''L/\xi_{\rm perc}}$. Hence, the final integral
converges, and one would expect it to be bounded by
$\lambda'L^{-1/\nu_{\rm perc}}$ [this would follow immediately, by
using the identity (\ref{ident}) once again, if we had more
information about the sign of the integrand in this identity]. The
following arguments provide a detailed support for this idea, and
indicate that this conservative bound is likely to be the precise
order of this correction term in most cases.

We will use the notion of finite-size scaling \cite{fb}, which
generalizes the scaling statements to finite size $L$. This
follows the form for conventional equilibrium phase transitions
(see especially Ref.\ \cite{bnpy,bzj}), which percolation closely
resembles (some rigorous results can be found in Ref.\
\cite{aizen,bcks}). We will briefly review the form of these
arguments, so as to include the cases $d>d_c$. While
$\overline{\cal N}_{\rm sing}(p,L)$ is analytic in $p$ for finite
$L$, we wish to identify a part (traditionally termed
``singular'') that in the vicinity of $p=p_c$ tends to $L^d Y_{\rm
sing}(p)$ as $L\to\infty$. This may be defined by subtracting the
nonsingular part of $Y(p)$:
\be%
\overline{\cal N}_{\rm sing}(p,L)= \overline{\cal N}(p,L)
-L^d(Y(p)-Y_{\rm sing}(p)), \ee%
which again will be used only in the region $p \simeq p_c$. Then
according to the theory of finite-size scaling for equilibrium
phase transitions, as $L\to\infty$, $\overline{\cal N}_{\rm
sing}(p,L)$
obeys the scaling form%
\be%
\overline{\cal N}_{\rm sing}(p,L)=n(tL^{y_t},uL^{y_u}), %
\ee where $t=p-p_c$, $u$ is an additional parameter (a coupling
constant) that in a field theoretic calculation \cite{hlhd} is
treated as independent, and $y_t$ and $y_u$ are universal scaling
dimensions (which depend on $d$). The scaling form is supposed to
hold for some finite function $n$ as $L\to\infty$ with the
arguments $tL^{y_t}$, $uL^{y_u}$ held fixed, and thus does apply
only for $p$ close to $p_c$. The correlation length, in an
infinite system, scales as $\xi_{\rm perc}\propto
|p-p_c|^{-\nu_{\rm perc}}$, where $\nu_{\rm perc}=1/y_t$. For
$d<d_c$, $u$ renormalizes to a fixed point value and can be
dropped (unless it is desired to find corrections to scaling). For
$d>d_c$, $u$ renormalizes towards zero ($y_u\propto d_c-d<0$), but
cannot be dropped as the free energy $n$ depends on it in a
singular fashion: %
\be%
n(x,z)=z^{p_1}n^*(xz^{p_2})\ee%
as $z\to0$. The authors of Ref.\ \cite{bnpy} showed that $p_1=0$,
and this should also hold for percolation. Then the mean number of
clusters takes the form%
\be%
\overline{\cal N}_{\rm sing}(p,L)= \left\{\begin{array}{ll}
n(tL^{y_t}) &\hbox{ for $d\leq d_c$,}\\
n^*(tL^{y_t^*}) &\hbox{ for $d\geq d_c$,}
\end{array}\right.,\ee%
in which $u^{p_2}$ has been absorbed into the non-universal scale
factors that accompany $t$. Here $y_t^*=y_t+p_2y_u$, and for
percolation the field-theoretic formulation leads to
$y_u=(6-d)/2$, $p_2=-2/3$, $y_t^*=d/3=y_t d/d_c$ \cite{jy} for
$d\geq6$. The implication of these scaling statements is that the
analytic background that has been subtracted has negligible
(exponentially small) $L$ dependence, even at $p_c$. The
finite-size scaling form given should be of order $L^d$ as
$L\to\infty$ with $t$ fixed, and must match $L^dY(p)_{\rm sing}$,
so we must have
\be%
\begin{array}{llll}
n(tL^{y_t}) &\sim C_\pm L^d|t|^{d/y_t}&\propto L^d \xi_{\rm
perc}^{-d}
&\hbox{ for $d\leq d_c$,}\\
n^*(tL^{y_t^*}) &\sim C_\pm L^d|t|^{d/y_t^*}&\propto L^d \xi_{\rm
perc} ^{-6}&\hbox{ for $d\geq d_c$,}
\end{array}\ee
for $|t|L^{y_t}$ (resp., $|t|L^{y_t^*}$) large, for both signs of
$t$. These scaling behaviors are consistent with the above forms
for $\alpha$, and $L^dY_{\rm sing}(p)$ itself satisfies the same
scaling behavior as $\overline{\cal N}_{\rm sing}(p,L)$,
$L^dY_{\rm sing}(p)=\overline{Y}(tL^{y_t})$
$(\overline{Y}(tL^{y_t^*})$ for $d\geq d_c$). For $d=d_c$ there
may be logarithmic corrections to these scaling forms, which we
will neglect.

Now the integral in eq.\ (\ref{formul}) contains only $\partial
\overline{\cal N}_{\rm sing}(p,L)/\partial p -$ $L^d \partial
Y_{\rm sing}(p)/\partial p$, and {\em for $d\leq d_c$ only} can be
rewritten using the scaling behavior in terms of $x=tL^{y_t}$
(we turn to the $d>d_c$ cases below):%
\be%
-L^{-y_t}\int_{-\infty}^{+\infty}dx\,x\left(\frac{d n(x)}{d
x}-\frac{d\overline{Y}(x)}{dx}\right).\ee %
The difference of derivatives is expected to behave as
$e^{-c''|x|^{1/y_t}}$ for some $d$-dependent constant $c''$ at
large $|x|$, because the leading error is due to correlations that
propagate around the system, and will involve the linear size
$L/\xi_{\rm perc}$. It follows that the integral converges, and we
have obtained%
\be%
\overline{\ell_{\rm OPT}}\sim \beta L^d -p_c+\lambda'L^\theta\ee%
as $L\to\infty$, with $\theta=-y_t$.

As an aside, we point out that $n(x)-\overline{Y}(x)$ cannot go to
zero at large positive $x$, but must approach $1$ as $p\to 1$. We
have pointed out that $\overline{\cal N}(p,L)-L^dY(p)$ approaches
$1$ as $p\to 1$; now we are arguing that this difference of order
1 exists all the way to the vicinity of $p=p_c$, and so the
integrand in eq.\ (\ref{ident}) behaves as a $\delta$-function
when $L\to\infty$. This effect is due to the ``giant'' percolation
cluster that occupies a positive fraction of vertices when $p >
p_c$. If we start at $p=1$, and decrease $p$, then edges are
removed at random. Some of these removals disconnect some vertices
from the giant cluster. However, the resulting value of
$L^{-d}\partial \overline{\cal N}(p,L)/\partial p$ has only small
finite size corrections, of relative order $e^{-c''L/\xi_{\rm
perc}}$. The giant cluster does not disappear until the critical
region is reached (where it cannot be distinguished from clusters
of size $\xi_{\rm perc}\simeq L$), and so $\overline{\cal
N}(p,L)-L^dY(p)$ remains close to 1 down to the same region.

A useful check on the arguments is provided by the $d=1$ case, in
which $\overline{\cal N}(p,L)=L(1-p) +p^L$,
$n(x)-\overline{Y}(x)=e^x$ ($x\leq 0$ for $d=1$). Thus%
\be%
\overline{\ell_{\rm OPT}}=L/2-1+L^{-1}+\ldots\ee%
in $d=1$ (higher terms are of order $L^{-2}$), that is
$\beta=1/2$, $\lambda'=1$. It is likely that $\lambda'>0$ for all
$d$.

For $d>d_c$, the use of the scaling forms with $n^*$ in place of
$n$ would lead to the final integral being of order $L^{-y_t^*}$.
This result is incorrect. The error is that while the scaling form
for $\overline{\cal N}_{\rm sing}(p,L)$ correctly describes the
{\em leading} behavior as $p\to p_c$, the integral we wish to
calculate contains the difference of derivatives, from which the
leading part has been subtracted. It turns out that there is a
subleading part of $\overline{\cal N}(p,L)$ that dominates this
subtracted form. Mean-field theory yields a nonanalytic
contribution to $\overline{\cal N}(p,L)$ that is precisely of the
form $L^d Y_{\rm sing}(p)$ near $p_c$. The leading correction to
$L^dY(p,L)$ due to Gaussian fluctuations at all wavevectors
(within a field-theoretic formulation) is $\sim C_\pm'L^d
|t|^{d\nu_{\rm perc}}$ (times $\ln |t|$ when $d$ is even), which
is smaller than $L^d Y_{\rm sing}(p)$ as $t\to 0$. [For
comparison, for $d>d_c$, the universal scaling function
$n^*(tL^{y_t^*})$ comes entirely from the ``zero-mode''
fluctuations \cite{bzj}.] However, when $L^d Y(p)$ is subtracted,
the leading singularity $L^d Y_{\rm sing}(p)$ is removed, and so
is $C_\pm'L^d |t|^{d\nu_{\rm perc}}$, but a finite-size correction
to the latter remains. This finite-size correction is of the form%
\ba %
\overline{\cal N}(p,L)-L^dY(p)&=&-\frac{1}{2}\left[\sum_{\bf
q}\ln(q^2+|t|)\right.\nonumber\\&&{}\left.\quad{}-L^d\int
\frac{d^dq}{(2\pi)^d}\ln(q^2+|t|)\right].
\ea The ultraviolet divergence in this expression is cut off on
the lattice; $q^2$ is replaced by a lattice expression that is
periodic over the Brillouin zone (to which the sum and integral
are restricted), and which reduces to $q^2$ at small $q$. The sum
is over wavevectors ${\bf q}=2\pi(n_1,\ldots,n_d)/L$, where $n_i$
are integers. Some numerical factors multiplying $t$ have been
neglected. One finds that $\overline{\cal N}(p,L)-L^dY(p)\propto
e^{-L|t|^{1/2}}$ as $L\to\infty$. This correction is significant
when $|t|<L^{-1/\nu_{\rm perc}}$. For $d>d_c$, the region
$|t|<L^{-1/\nu_{\rm perc}}$ is much larger than $|t|>L^{-y_t^*}$,
within which the other effects are important. In the wider region,
the Gaussian fluctuations dominate, as the interaction term $u$ is
weak (and perturbation theory is infrared convergent for $d>d_c$).
The contribution of the giant cluster also is significant over the
same window. Then $\partial \overline{\cal N}(p,L)/\partial p -
L^d \partial Y(p)/\partial p$ possesses a scaling limit that is a
function of $tL^{y_t}$ only, where $y_t=2$ in this case. Hence the
rescaling argument in this case produces $\lambda' L^{-1/\nu_{\rm
perc}}$ also. There are also other corrections for $d>d_c$,
including an effective finite-size shift in the value of $p_c$, of
order $L^{-(d-4)}$, which is smaller than the width of the
critical region $|p-p_c|\propto L^{-y_t}$. This shift contributes
an amount of order $L^{-(d-4)}$ to $\overline{\ell_{\rm OPT}}$,
smaller than $L^\theta$. For $d\leq d_c$, all fluctuation effects
are of similar order as the leading mean-field term, and have to
be resummed using the renormalization group; they contribute to
the same universal scaling functions $n$ and $\overline{Y}$, and
the present arguments for $d>d_c$ do not apply there.

The generalization to finite sizes with periodic boundary
conditions, but for a cuboid of general aspect ratio (held fixed
as $L\to\infty$) in place of the hypercube, is straightforward.
Another generalization is to a long cylinder, of length $L$, and
hypercubic with periodic boundary conditions with period $W$ in
the $d-1$ transverse dimensions. In this case, the mean optimum
cost per unit length tends to a $W$- (and $d$-) dependent limit as
$L\to\infty$, and by similar arguments (using methods from Ref.\
\cite{bzj} for this geometry) this behaves as
\be%
\lim_{L\to\infty}\overline{\ell_{\rm OPT}}/L\sim \beta W^{d-1}+
\lambda''W^{\theta-1},\ee %
with the same exponent $\theta$, as $W\to\infty$.

Finally, the application of similar ideas to the simplest model
MST on the complete graph with $N=|V|$ vertices, to obtain
finite-$N$ corrections to the result of Ref.\ \cite{frieze},
should give (using an analysis like Ref.\ \cite{bzj}, and similar
to the $L^{-y_t^*}=N^{-1/3}$ for $d>d_c$ that we argued is
incorrect in the finite-$d$ lattice case, but should be
correct here)%
\be \overline{\ell_{\rm OPT}}\sim\zeta(3) -1/N+\lambda'''N^{-4/3}
\ee [we note that the percolation threshold is $p_c=1/N$ (see e.g.
Ref.\ \cite{bollo}), and all terms are smaller by $1/N$ than in
the lattice cases].


\subsection{Effect on the mean cost of a change of boundary condition}
\label{sec:boun}

In this subsection, we consider the long-cylinder geometry,
described in the introduction and at the end of section
\ref{sec:fin}. We consider the effect of a change in boundary
condition, that is imposed by demanding that the minimum cost
spanning tree have $k$ distinct branches crossing from one end to
the other, instead of the one that is typical for the usual MST.
We call the minimum cost for this constrained spanning tree
$\ell_{\rm OPT}(k)$. Thus, outside of the end regions of the
cylinder, there are (at least) $k$ trees, forming a spanning
forest of minimum cost. This type of change of boundary condition
could be handled by the Hamiltonian methods described in Ref.\
\cite{bzj}, if we had a direct field theory for the MST problem.
This would lead us to expect the change in optimum cost per unit
length to scale the same way as the finite $W$ correction to the
optimum, that is as $W^{\theta-1}$. This expectation is correct,
but as such a formulation is not presently available, we will turn
to a different approach, which produces an upper bound, and which
can also be applied to other combinatorial optimization problems.

The idea is to begin with the MST on the long cylinder without the
additional constraint, and now modify it so as to grow $k-1$
additional disconnected trees that extend from one end to the
other. This must increase the total cost, and we estimate the
resulting increase, thus producing an upper bound on this change.

It is useful to give two versions of this procedure; the first
version is simple and produces a rather conservative bound, while
the second, more refined upper bound is tighter. When expressed in
terms of an exponent $\theta$, which should be the same as the
other $\theta$s in this paper, the first says that $\theta\leq0$,
and the second that $\theta\leq -1/\nu_{\rm perc}$. The second
bound presumably cannot be tightened further in most cases.

We begin with some definitions for the MST on a long cylinder.
There is a path on the tree from one end of the cylinder to the
other, which with probability approaching 1 as $L/W\to\infty$ is
unique outside the end regions (of length of order $W$). As the
end regions are unimportant, this path is essentially unique, and
we will refer to it as the {\em trunk} of the tree. The remainder
of the tree consists of side-branches, which are trees rooted on
the trunk; the side-branches presumably have linear size of order
$W$ or less. The basic procedure, which we describe for $k=2$ as
the generalization to $k>2$ is simple, is to modify the tree in a
sequence of steps so as to grow a second tree, distinct from what
remains of the first one except in the end regions, that possesses
a trunk extending from one end to the other of the cylinder. This
is done by beginning at one end of the cylinder, and cutting off
parts of side-branches (by removing an edge) from the original MST
and joining them to the new tree. Each side-branch of the original
tree that is cut must be adjacent to the new, growing tree so that
it can be reattached to it, by including an edge that was not part
of the original MST. We end up with two disjoint trees, which
together span the vertices, one of which has the same trunk as the
original MST. The side-branches, and the cutting and attaching
edges, are selected so as to minimize the increase in cost of the
final $k$ trees relative to the MST.

In the first, simple procedure, at each step we look for a
side-branch attached to the trunk of the original MST that is
adjacent to the growing tree, and which extends in the growth
direction. This will typically be of size $W$, and will touch the
growing tree at a distance of order $W$ from the trunk of the MST.
The cut is made at an arbitrary point between the re-attachment
point (which is also chosen arbitrarily) and the trunk. The
growing tree thus grows by order $W$ towards the target end. The
number of steps required will be of order $L/W$, and each
increases the cost by order one, so the change in cost is of order
$L/W$, and the pair of trees constructed provides an upper bound
on the true minimal increase in cost relative to the MST. For the
general $k$-tree version, $k-1$ additional trees can be grown in
parallel, and each step makes progress by $W/k^{1/(d-1)}$, similar
to arguments in Ref.\ \cite{ABNW}. The total change in cost is
then roughly of order $k^{d/(d-1)}L/W$.

In the second, improved version, we will recognize that the
selection of edges to cut and to add can be optimized to
significantly reduce the increase in cost per step. In fact, the
edges that will be moved will again be ``critical edges'', here
meaning those with cost within $W^\theta$ of $\tilde{\ell}_c$.

If the greedy algorithm is applied to any one of our models on the
cylinder, then we can run it up to a value of
$\tilde{\ell}<\tilde{\ell}_c$ such that $\xi_{\rm perc}<W$, say
$\xi_{\rm perc}=W/10$. If $W$ is large, this means
$\tilde{\ell}=\tilde{\ell}_c$ minus of order $W^{-1/\nu_{\rm
perc}}$. At this stage, there are many clusters of size $\xi_{\rm
perc}$, but it is rare for a cluster to percolate around the
``circumference'' $W\ll L$ of the cylinder (the probability in a
length of order $W$ along the cylinder is of order
$e^{-cW/\xi_{\rm perc}}$). We will ignore these exceptional cases,
for now, and return to this oversimplification later. Now we
continue the greedy algorithm up to $\tilde{\ell}=\tilde{\ell}_c$
plus of order $W^{-1/\nu_{\rm perc}}$. There will now be many
large clusters that have size $\gg W$ along the long direction of
the cylinder, and which together occupy a positive fraction of
vertices as $L\to\infty$. However, we cannot guarantee that there
is a single giant cluster that percolates the full length of the
cylinder with probability one. There is always a nonzero
probability that the cluster is broken somewhere, even though this
probability may be exponentially small in $W$. (For finite $W$, on
length scales $>W$ the problem maps onto an effectively one
dimensional percolation problem, in which $p_c=1$.) In fact, if
above $\tilde{\ell}_c$ the correlation length is $\xi_{\rm perc}$,
then the probability per unit length of a break in the cluster is
of order $e^{-(W/\xi_{\rm perc})^{d-1}}$ when $W\gg \xi_{\rm
perc}$. This will not affect the argument, and we may continue as
if there is a giant cluster and a path on the corresponding tree
that runs from one end to the other (this path will be the trunk
of the MST when the greedy algorithm is finished). After giving
the argument under this simplifying but incorrect assumption, we
will return to and correct for this oversimplification also.

We can choose $\tilde{\ell}-\tilde{\ell}_c$ large enough so that
there are actually two (or more generally, $k$) paths from end of
the cylinder to the other on the giant percolation cluster, which
have no edges or vertices in common with one another. We will
assume that the paths are separated by of order $W/2$ (or
$W/k^{1/(d-1)}$ for $k\geq 2$) along almost all of their length
(again, this may be an oversimplification, but should not affect
the scaling). Now on the corresponding tree (which is a subset of
the edges of the cluster), there is only one path (or trunk)
running from one end to the other. Take the trunk as one of the
two disjoint paths on the cluster. The second path runs along the
tree, but suffers many breaks at edges that are part of the
cluster but not of the tree. The parts of the path that are edges
on the tree lie on side-branches off the trunk, and typically some
of the edges that connect this path to the trunk were not present
at $\tilde{\ell}=\tilde{\ell}_c - W^{-1/\nu_{\rm perc}}$. If we
take this tree and remove one of these edges, and replace them
with the edges on the cluster that complete the second path, then
we have satisfied the constraint on the tree, and the remainder of
edges of the MST can be added to these two trees without producing
any cycles. Thus we have constructed two trees that together span
the vertices, with two disjoint paths running from one end to the
other, at an increase in cost of order $W^\theta$ per length $W$,
with $\theta=-1/\nu_{\rm perc}$. Note that, as in the simple
version of the argument, we expect that only of order one edge
(i.e.\ a fixed number as $W\to\infty$) must be moved per length
$W$ along the cylinder in order to construct the second path.

The existence of breaks on the trunk of the MST when the algorithm
stops at $\tilde{\ell}$ when the tree is not spanning does not
affect the above argument (after all, we can easily ensure that
the second path is disjoint from the whole trunk of the MST). The
second path that is constructed will also have breaks on it. These
can be filled as $\tilde{\ell}$ increases further. They become
exponentially rare when $W\gg \xi_{\rm perc}$, so that the
increase in cost for moving edges to construct the second trunk
will converge to $W^\theta$ per length $W$, as claimed. Similarly,
the clusters that encircle one (or more) of the periodic
directions of the cylinder when
$\tilde{\ell}=\tilde{\ell}_c-W^{-1/\nu}$ are avoided if we go to
even smaller $\tilde{\ell}$. The total contribution of these
events will converge and still scale as claimed.

The refined version of the argument thus suggests that the change
in cost per unit length is bounded by, and most likely actually of
order of, $W^{\theta-1}$ times $k$- and $d$- dependent factors, as
claimed earlier,%
\be%
\lim_{L\to\infty}\frac{\overline{\ell_{\rm
OPT}(k)}-\overline{\ell_{\rm
OPT}}}{L}\sim\lambda'_kW^{\theta-1}\ee%
as $W\to\infty$.


\subsection{Scaling at finite size and positive temperature}

Our final topic for MSTs will be the combined effects of small
positive temperature and finite size. We again assume the system
is a hypercube of side $L$ with periodic boundary conditions. We
consider the mean free energy $\overline{F}$, where $F=-T\ln Z$,
and subtract the non-singular part, as in the theory of
finite-size scaling for critical phenomena at non-zero temperature
discussed in section \ref{sec:fin}. The non-singular part takes
the form $L^d(\beta +a''T^2+a_1'T^3)-\tilde{\ell}_c$ (there is a
possibility of terms of order $T^2L^0$ also). For the singular
part $\overline{F}_{\rm sing}$ we have the scaling form%
\be%
\overline{F}_{\rm sing}(T,L)=T{\cal F}(TL^{y_T})\label{Fscal}\ee%
for $d\leq d_c=6$. Here the exponent $y_T$, the scaling dimension
of $T$, will turn out to be $y_T=-\theta=1/\nu$. The factor of $T$
occurs because (as mentioned in section \ref{posT}) it is $F/T$
that scales similarly to $F$ in the nonzero temperature critical
phenomena case. The scaling function ${\cal F}(x)$ is a function
of the natural scaling combination $x=TL^{1/\nu}$, and scaling is
supposed to hold as $T\to0$ (and $L\to\infty$) with $x$ fixed. It
has the limiting
behavior%
\be%
{\cal F}(x)\propto\left\{\begin{array}{ll}
           x^{d/y_T},&\hbox{ as $x\to\infty$,}\\
            x^{-1},&\hbox{ as $x\to0$.}
                                    \end{array}\right.\ee%
These two limits reproduce the results of the previous sections,
in the two limits $L\to\infty$ with $T$ fixed, and $T\to0$ with
$L$ fixed, provided $y_T=-\theta$. We emphasize that at finite
$L$, the explicit average over the disorder is required, as $F$
itself is subject to fluctuations in the scaling limit. It should
be possible to describe the statistics of the fluctuations in the
singular part of $F$ by scaling forms with the same exponents,
also. Note that the non-singular part we subtracted included the
non-singular subleading (in terms of $1/L$) term
$-\tilde{\ell}_c$, so that the scaling function exposes the parts
with non-trivial exponents, such as $d\nu$ or $\theta$ in the two
limits.

For $d>d_c$, we find some difficulty in obtaining a convincing
scaling form that reproduces the limits in previous sections. This
is due to hyperscaling being violated in the positive temperature
results ($\overline{F}(T,L)_{\rm sing}\propto L^d T^4$), but not
in the finite size results ($\overline{F}(0,L)-\beta L^d
+\tilde{\ell}_c\propto L^{-1/\nu}$). Possibly the problem is due
to the singular part of the positive temperature result not being
unambiguously distinguishable from the analytic behavior, as we
have already discussed. Likewise, the finite-size contribution at
$T=0$ is due to long-range correlation effects, but is an integer
power of $L$ ($L^{-2}$). It cannot in principle be distinguished
from a nonsingular part of the same order. Even though we did not
find such a terms, we did have to subtract a term of order $L^0$.
As we saw in the case of percolation, above the critical dimension
there may be contributions to the free energy that scale in
distinct ways. We suspect that we must write the general form as
\be%
\overline{F}_{\rm sing}(T,L)=T{\cal F}_1(TL^{y_T})
+T{\cal F}_2(TL^{y_T^*})\label{Fscald>}\ee%
The functions in this expression have the limiting behavior
\be%
{\cal F}_1(x)\propto\left\{\begin{array}{ll}
           O(x^{d/y_T}),&\hbox{ as $x\to\infty$,}\\
            x^{-1},&\hbox{ as $x\to0$.}
                                    \end{array}\right.\ee%
and $y_T=-\theta=1/\nu=2$ for $d\geq d_c=6$, while
\be%
{\cal F}_2(x)\propto\left\{\begin{array}{ll}
           x^{d/y_T^*},&\hbox{ as $x\to\infty$,}\\
            O(x^{-1}),&\hbox{ as $x\to0$.}
                                    \end{array}\right.\ee%
where $y_T^*=y_Td/d_c=d/3$ for $d\geq 6$. (Here, as usual,
$X=O(Y)$ as $Z\to\infty$ means $|X/Y|$ is bounded as
$Z\to\infty$.) Each of the two previous scaling limits is
reproduced by one of these two functions, while the other is
smaller in that limit. In the two scaling limits of this
paragraph, in each of which some combination of $T$ and $L$ is
held fixed in the limit, one of the two functions dominates (and
takes a limit form calculated in one of the previous sections),
while the other (the one that is a function of the combination
held fixed) describes subleading corrections. A more complete
study of this issue would be of interest.


\section{Other Optimization Problems}\label{sec:oth}

In this section we consider possible extensions of the results to
other combinatorial optimization problems that have a geometric
flavor.

The first one to mention is the minimum Steiner tree (MStT)
problem \cite{steele,law}. In its Euclidean version, there are $N$
``mandatory'' points marked in a region $\Lambda$, and we must
find a tree that visits all of them with minimum total Euclidean
length for its edges, similar to the Euclidean MST, but now it is
allowed to have vertices of the tree that are not mandatory. There
is also a version on a graph, in which a subset of the vertices
are mandatory, costs are assigned to the edges, and a minimum cost
tree must be found that visits all the mandatory points. While the
MST can be solved in a time polynomial in $|V|$ (using e.g.\ the
greedy algorithm \cite{papst}), the MStT is NP-hard (i.e.\ the
decision version, asking the question whether there exists a
Steiner tree with cost less than some given value, is NP-complete
\cite{papa}) and presumably cannot be solved in polynomial time.
Both optimization problems produce a tree that (in the random
version of the problem) fills space on large scales (with high
probability), thus similar connectivity and boundary-condition
properties can be defined. It is plausible that the scaling
dimensions for the MStT are the same as for the MST, including
$\theta$ and $\nu$ as defined in this paper. This would be
analogous to universality arguments in statistical mechanics
problems such as Ising spin problems, in which universality
classes can be distinguished on the basis of the locality and
symmetry of the Hamiltonian and of the type of disorder involved.
For geometric problems of the type considered here, there are no
local order parameters (analogous to spins), but topological
properties such as the connectivity we have used should take their
place.

We can consider coarse-graining methods, which we here describe
schematically. Coarse graining, or renormalization, is designed to
preserve the properties that define universality classes. If we
consider the points within a window of size $W$ within the sample,
then the tree passes through its boundary at one or more points
(with probability approaching 1 as $W$ increases). Only the fact
that each of these is or is not connected through the interior of
the window to each other such point (for the given window) is
relevant to the tree outside. Thus minimization of the cost over
the interior can be performed for each such boundary condition. If
the system is partitioned into such windows of equal size, then
patching together the windows subsequently, one can minimize the
total cost in stages that are performed locally, at the cost of
storing a large amount of information about the results for
different boundary conditions. The information that needs to be
stored is reduced by coarse-graining, that is assuming that fine
details of the structure will not be important. In particular, in
low dimensions (less than eight \cite{ns}) there will typically be
only a finite number of large (size of order $W$) trees visible
within each window, even for large windows. The reduced objects
can be represented as trees, but with a lower density of vertices.
These are the usual ideas of the renormalization group, applied to
geometric objects. In general, the cost for given connections
within a window will depend on the connections in a complicated
way, and cannot be expressed simply as a sum over some
``occupied'' edges. One property that should be maintained as
coarse-graining proceeds is that if two disconnected portions are
connected, the cost will increase. Thus, the simple form of the
cost for the MST, and the less simple (in terms of the mandatory
vertices) form for the MStT, are just two examples, and all models
will become more complicated under coarse-graining anyway. It is
then likely that the universality classes (one for each dimension
$d$) in which all the (short-range correlated, $d$-dimensional)
MST problems lie are actually larger and contain some more general
tree-optimization problems. Hence it is not at all implausible
that the MST and MStT are in the same universality class for each
$d$.

There are also other popular problems, such as the traveling
salesman problem (TSP), and minimum weighted matching. The scaling
forms for various quantities given in previous sections should
also apply to these (in their $d$-dimensional version), though the
universal numbers, including the exponents and critical
dimensions, may be different. For the TSP, we can define $\theta$
from the finite-size correction to the total cost, say for
periodic boundary conditions on a hypercube, as
$\overline{\ell_{\rm OPT}}=\beta L^d+\lambda'L^\theta+\ldots$. For
the TSP, Rhee \cite{Rhee} raised the question (for $d=2$) of
whether for periodic boundary conditions, in our notation,
$(\overline{\ell_{\rm OPT}}-\beta L^2)/L\to 0$ as $L\to\infty$.
Our answer to this question would be affirmative. We note that the
order one term in $\overline{\ell_{\rm OPT}}$ for MST with
periodic boundary conditions can be traced back to the fact that
$\ell$ is a sum of $|V|-1=L^d-1$ terms, not $|V|=L^d$. For the
TSP, $\ell$ is a sum of exactly $|V|$ terms. Alternatively, we can
define $\theta$ by considering the change in cost when the tour is
required to travel from one end of a cylinder to the other $k$
times, as in Ref.\ \cite{jrs}.

For the TSP at nonzero temperature, no phase transition is found
in mean-field theory \cite{mezpatsp}, and so we expect none in any
dimension $d$. The high-temperature limit of TSP is a sum over all
tours of the graph, so could be called ``uniform Hamiltonian
cycles'', but this is also essentially what is called dense
polymers (self-avoiding walks constricted in volume). However, we
should caution that uniform Hamiltonian cycles on some
two-dimensional lattices are known to be in different universality
classes from the more generic dense polymers; these are called
fully-packed loop models. In dense polymers, weak disorder is an
irrelevant perturbation, so it is reasonable to imagine that the
renormalization group can flow to the high-temperature fixed
point. Given the absence of a transition at finite non-zero $T$,
we expect that any positive temperature is relevant, and so that
$\theta\leq0$. Assuming that $\theta\leq0$, there will be a
crossover length $\xi\propto T^{-\nu}$ that diverges as $T\to0$,
with again the scaling relation $\nu=-1/\theta$. We can also try
to bound $\theta$ as in section \ref{sec:boun}. In the absence of
detailed information, we can still use an argument similar to the
simple bound given there. In particular, in two dimensions, the
tour is equivalent to the boundary of a tree, so that the argument
is really the same, and we conclude again that $\theta\leq0$. In
Ref.\ \cite{jrs}, it was assumed that $\theta=0$ for $d=2$, and
some support for this was found numerically.

More speculatively, since the two-dimensional TSP is equivalent to
minimizing a complicated but local cost function for a tree, the
type of coarse-graining arguments outlined above suggest that
$\theta_{\rm TSP}(d=2)=\theta_{\rm MST}(d=2)=-3/4$ (and that other
corresponding exponents also are equal, as suggested in Ref.\
\cite{jrs}). Even if this suggestion is correct, the universality
classes for TSP in dimensions $d>2$ do not have to join smoothly
with the MST class at $d=2$. There are actually (at least) two
probability measures for space-filling curves (or dense polymers)
in $d=2$, depending on whether they are strictly non-intersecting,
or self-intersections are discouraged but not forbidden
\cite{jrs1}. Whether or not TSP is in the same universality class
as dense polymers in any dimension, or for any subset of its
properties, a similar topological distinction probably holds for
TSP \cite{jrs}. A two-dimensional version of the TSP that allows
the curves to cross can be obtained using a tour in a
three-dimensional slab of small thickness in one direction, that
is large in the two orthogonal directions. On large scales, this
problem is effectively two dimensional, and the optimum tour
projected into these two dimensions will intersect itself. Such
problems will define a distinct universality class of TSPs from
the usual planar (non-self-intersecting) one. It will be the
natural continuation of the TSP universality class for $d>2$ to
$d=2$, as in the case of dense polymers \cite{jrs1,jrs}. The
suggestion in Ref.\ \cite{jrs} that the TSP is in the universality
class of dense polymers for $d\geq 2$ (where the $d=2$ case means
the version with intersections) implies that the critical
dimension is $2$, at least for the geometric correlation
properties (that is, $d=2$ is analogous to $d=8$ for MSTs
\cite{ns}). It would be interesting to use the mean-field approach
\cite{mezpatsp} in finite dimensions to calculate a mean-field
value of $\theta$ for the TSP for sufficiently high $d$, and to
find the value of $d_c$ for the TSP.

In an interesting paper, Moore \cite{moore} applied the idea of
$\theta$ (which he called $y$) to combinatorial optimization
problems. He argued that for the TSP, $\theta=1$ for all
dimensions $d$. His argument was based on the analysis of the
relative error in a partitioning algorithm by Karp \cite{karp}.
Inspection of this analysis shows that the relative error is
related to the first boundary term in an expansion for a hypercube
with free boundaries, $\overline{\ell_{\rm OPT}}\sim \beta
L^d+\beta_1 L^{d-1}+\ldots$ ($\beta_1$ has been shown to be
positive \cite{Rhee}). If a large system is partitioned into such
cubes, which are solved separately, then there will be errors of
this form for each cube \cite{karp}, which would be absent in a
better scheme. Further, as we have seen, the boundary terms for
the whole system do {\em not} scale with exponent $\theta$.
Accordingly, we do not believe that this is a valid determination
of the value of $\theta$ for the TSP.

A perfect matching is by definition a subgraph of $G$ that
includes all the vertices of $G$, such that every vertex is on
exactly one edge of the matching. In minimum weighted matching,
one must find a perfect matching such the cost, which is the sum
of the costs of the ``occupied'' edges (those on the matching), is
minimized \cite{papst,steele,law}. The case in which $G$ is
bipartite (there are two sets of vertices $U$ and $V$, with
$|U|=|V|=N$, and only edges that connect a member of $U$ to one of
$V$) is a little easier to solve, and is also known as the
assignment problem. The Euclidean bipartite minimum matching
problem (which is also known as two-sample matching), in which the
vertices in $U\cup V$ are distributed, for example, independently
and uniformly over a domain such as $[0,L]^d$ (with $N/L^d=1$) has
the curious property (as quoted in Ref.\ \cite{steele}) that, at
least for two dimensions, the mean optimum cost is of order
$L^2(\ln L)^{1/2}$. This is not the case for the unrestricted
(non-bipartite) Euclidean problem \cite{steele}. Minimum weight
matching occurs (though not with Euclidean distance as the cost)
in finding the ground state of an Ising spin glass in two
dimensions, with free boundary conditions, and also in other
physical problems. Leaving aside cases like the two-dimensional
Euclidean bipartite one that may require special treatment, we
again argue that $\theta\leq 0$, on the basis of the absence of a
transition in mean-field theory \cite{mezpamm}. There is a similar
picture of positive temperature causing a flow to the ``uniform
matchings'' problem, also known as ``dimer packing''; in this, the
high temperature limit of the partition function, the sum is over
all matchings with equal Boltzmann-Gibbs weight.

It should not be imagined that $\theta<0$ in all combinatorial
optimization problems, even in those that can be solved in
polynomial computation time. The shortest-path problem (for two
given vertices separated by distance $L$, find the path between
them of lowest total cost, where non-negative random costs are
assigned independently to each edge of a lattice) is equivalent to
the directed polymer problem (see especially Ref.\ \cite{fh}). The
variations in the cost of the optimal path scale as $L^\theta$,
with $\theta>0$ for all dimensions $d\geq 1$, and $\theta=1/3$ in
two dimensions. If the cost is viewed as ``time'', then
shortest-path becomes first-passage percolation. Other
generalizations of shortest-path have been considered in
statistical physics, including that in which the directed path is
replaced by a $d$-dimensional surface (e.g.\ a domain wall) in
$d+1$-dimensional space, each point of the surface has a unique
projection to the $x_{d+1}=0$ coordinate hyperplane, and the cost
(or energy) is the sum of random costs assigned to faces of the
lattice occupied by the surface \cite{gmf}. We will assume here
that the projection of the surface to $x_{d+1}=0$ is a
$d$-dimensional hypercube of side $L$, and that the boundary of
the surface is fixed in the $x_{d+1}=0$ hyperplane. For $d\geq
d_c=4$, $\theta$ takes on the mean-field--like value $d-2$, when
$-\theta$ is defined as the scaling dimension for temperature
\cite{gmf}. However, the leading finite-size correction to $\beta
L^d$ in the mean optimum cost (ground state energy) involves the
disorder, which is irrelevant for $d\geq 4$, and hence the
correction term is $\lambda'L^{d-2}/L^{d-4}=\lambda' L^2$.


\section{Conclusion}

The central results of this paper concern the behavior of the
correlation length $\xi\propto T^{-\nu}$ as $T\to0$, and the
finite size correction to the optimum cost $\propto L^{\theta}$,
with the scaling relation $\nu=-1/\theta$. We find that
$\nu=\nu_{\rm perc}$, the correlation length exponent in classical
percolation, for all dimensions $d$. This result rests on the
identification of the ``critical edges'' that have cost close to
the percolation threshold, as these edges connect the tree over
large scales, and can be replaced by one another at low change in
cost (of order $T$ or $L^{-1/\nu}$ per edge for the positive
temperature, and finite size situations, respectively). Although
it is sometimes said that there is no phase transition behavior in
optimization, the results presented here can be understood as a
transition occurring right at $T=0$.

We used Kruskal's greedy algorithm in many of the arguments, but
the results we obtain are about the MST (or near optimal,
thermally excited trees), and do not depend on the algorithm used.
Thus this is not an ``analysis of an algorithm'' in a traditional
sense. There may still be more to be learned by using other
algorithms. It would be interesting to analyze other problems that
possess polynomial-time algorithms (notably, minimum matching) in
a similar manner.

The discussion of universality classes, and our suggestions (see
also Ref.\ \cite{jrs}) that minimum spanning tree, minimum Steiner
tree, and even two-dimensional traveling salesman problem may be
in the same universality class, serves to illustrate that the
universal scaling properties discussed in this paper may have very
little to do with the computational complexity issues of P versus
NP \cite{papa}, which seem to depend entirely on details of the
definition of the optimization problem at short length scales
(some related observations are made in Ref.\ \cite{cmoore}).
Possibly this is due to the difference between the average-case
behavior that is analyzed here and related to universality
classes, and the worst-case computational complexity characterized
by P or NP. On the other hand, the scaling properties may be very
useful for understanding the effectiveness of algorithmic
techniques (such as local search and randomized algorithms) and
approximation schemes, when they are applied to hard random
problems in $d$ dimensions.

I am grateful to S. Girvin, P. Jones, R. Kannan, S. Sachdev, H.
Saleur, and D. Wilson for helpful discussions, references, and
correspondence. This work was supported by the NSF under grant
no.\ DMR-02-42949.




\begin{references}
\bibitem{limit}      Throughout this paper, we use notation $X\sim Y$ as $Z\to\infty$
                     in the strict sense: $\lim_{Z\to \infty}X/Y =
                     1$. Likewise, $X\sim Y_1+Y_2+\ldots$ as
                     $Z\to\infty$ means that $\lim_{Z\to \infty}X/Y_1 =
                     1$, $\lim_{Z\to \infty}(X-Y_1)/Y_2 =
                     1$, and so on.
\bibitem{papst1}    One way to show this is as follows: Compare the
                    given tree with the MST edge by edge, starting with
                    the least-costly edge on the MST. If the tree is
                    not the MST, we can find the least-costly edge
                    (say, $\langle ij\rangle$) on the MST that is
                    not on the given tree. Adding this to the given
                    tree creates a cycle. This cycle must include
                    at least one edge not on the MST; choose the
                    least costly one (suppose it is $\langle kl\rangle$).
                    This edge $\langle kl\rangle$ has $\ell_{kl}\geq \ell_{ij}$
                    (otherwise there is a contradiction with the greedy
                    algorithm). Hence it can be moved from $\langle kl \rangle$
                    to $\langle ij \rangle$ to (with probability one,
                    strictly) lower the cost.
\end{references}
\end{document}